\documentclass[apj, numberedappendix]{emulateapj}

\usepackage{amssymb}
\usepackage{natbib}

\newcommand{\eg}{\textit{e.g.}}
\newcommand{\ie}{\textit{ie.}}
\newcommand{\viz}{\textit{viz.}}
\newcommand{\cf}{\textit{cf.}}
\newcommand{\HI}{H{\footnotesize I}}

\newcommand{\HeI}{He{\footnotesize I}}
\newcommand{\htwo}{H$_2$}
\newcommand{\hz}{$\mathrm{H}^0$}
\newcommand{\hp}{$\mathrm{H}^+$}
\newcommand{\hm}{$\mathrm{H}^-$}
\newcommand{\htz}{$\mathrm{H}_2^0$}
\newcommand{\htp}{$\mathrm{H}_2^+$}
\newcommand{\hez}{$\mathrm{He}^0$}
\newcommand{\hep}{$\mathrm{He}^+$}
\newcommand{\hepp}{$\mathrm{He}^{++}$}
\newcommand{\eminus}{$\mathrm{e}^-$}
\newcommand{\goesto}{$\rightarrow$}
\newcommand{\subh}{_{\mathrm{H}}}
\newcommand{\subhtwo}{_{\mathrm{H}_2}}
\newcommand{\subHI}{_{\mathrm{H{I}}}}
\newcommand{\txth}{\mathrm{H}}

\newcommand{\mtwo}{M_{200}}

\newcommand{\subcrit}{_{\mathrm{crit}}}
\newcommand{\emm}{{\cal M}}
\renewcommand{\vee}{{\cal V}}
\renewcommand{\sun}{$_{\odot}$}
\newcommand{\Rmin}{{\cal R}_{\mathrm{min}}}

\begin{document}
\author{Edward N Taylor and R L Webster}
\affil{School of Physics, the University of Melbourne}

\shortauthors{E N Taylor and R L Webster}
\email{ent@strw.leidenuniv.nl}

\title{On Star Formation and the Non-Existence of Dark Galaxies}
\shorttitle{On Star Formation and Dark Galaxies}

\slugcomment{Accepted for publication in \apj, August 16, 2005}


\begin{abstract}
We investigate whether a baryonic dark galaxy or `galaxy without
stars' could persist indefinitely in the local universe, while
remaining stable against star formation.  To this end, a simple model
has been constructed to determine the equilibrium distribution and
composition of a gaseous protogalactic disk. Specifically, we
determine the amount of gas that will transit to a Toomre unstable
cold phase via the \htwo\ cooling channel in the presence of a
UV---X-ray cosmic background radiation field.

All but one of the models are predicted to become unstable to star
formation: we find that, in the absence of an internal radiation
field, the majority of gas will become Toomre unstable in all putative
dark galaxies with baryonic masses greater than $10^9$ M\sun , and in
at least half of those greater than $10^6$ M\sun .  Moreover, we find
that all our model objects would be detectable via \HI\ line emission,
even in the case that star formation is potentially avoided.  These
results are consistent with the non-detection of isolated
extragalactic \HI\ clouds with no optical counterpart (galaxies
without stars) by the \HI\ Parkes All-Sky Survey.

Additionally, where star formation is predicted to occur, we determine
the minimum interstellar radiation field required to restore
gravothermal stability, which we then relate to a minimum global star
formation rate.  This leads to the prediction of a previously
undocumented relation between \HI\ mass and star formation rate that
is observed for a wide variety of dwarf galaxies in the \HI\ mass
range $10^8$---10$^{10}$ M\sun.  The existence of such a relation
strongly supports the notion that the well observed population of
dwarf galaxies represent the minimum rates of self-regulating star
formation in the universe.
\end{abstract}

\section{Introduction}

There has been long-standing speculation that optically selected
galaxy surveys might be missing a population of very low surface
brightness (LSB) objects \citep{Disney, DisneyPhillips}.  It has
recently become possible to address this question observationally: the
\HI\ Parkes All-Sky Survey (\HI PASS) has now mapped the entire
Southern sky at $\sim$ 21 cm.  Since the \HI PASS catalogue
\citep{MeyerEtAl} represents an optically unbiased sample of neutral,
atomic hydrogen (\HI ) in the local ($z \lesssim 0.04$) universe, it
is ideal for the detection of \HI -rich, baryonic dark galaxies;
`galaxies without stars'.

The result of the \HI PASS search for dark galaxies is emphatic and
surprising.  With few possible exceptions [\eg\ \citet{KilbornEtAl,
RyderEtAl}; see also \citet{MinchinEtAl}], no isolated, extragalactic
\HI\ clouds were discovered for which an optical counterpart could not
be identified \citep{Doyle}; it seems there are no dark galaxies.

Prior to \HI PASS, the discovery of a number of `dim' galaxies ---
``galaxies where luminous matter is only a minor component of the
total galaxy mass'' \citep{CarignanFreeman} --- had led to the
expectation that star formation (SF) might be suppressed completely 
within some galaxies.  The first such examples were dwarf irregular
galaxies with extended \HI\ envelopes; DDO 154 serves as the archetype
of this class of gas-rich LSB dwarfs \citep{KrummBurstein}.  At the
other end of the mass scale, the serendipitous discovery of Malin 1
\citep{BothunEtAl, ImpeyBothun}, a very massive, \hbox{\HI -dominated}
LSB disk galaxy, provided an example of the class of ``crouching
giants'' predicted by \citet{Disney}.

A common feature of Malin 1 and DDO 154 is their high \HI\ mass:light
ratios: $M\subHI / L_B \sim 5$ [compared to $M\subHI / L_B \sim
0.1$---1 for `normal' \HI -detected LSB galaxies \citep{FisherTully,
Waugh}].  In addition to the observational problem of determining
their abundance --- which \HI PASS has begun to address --- galaxies like
DDO 154 and Malin 1 pose the following theoretical problem: ``Why was
there so little gas processed into stars'' in these systems
\citep{CarignanBeaulieu}?

Are these objects exceptional in the way in which they are presently
forming stars?  How far might this `dim' population extend toward
complete darkness?  And if dark galaxies can exist, is there some
mechanism to prevent their detection by \HI PASS?

In this paper, we consider the viability of dark galaxies in the local
universe; that is whether a dark galaxy, having formed, can remain
dark, or whether it will inevitably `light up'.  A simple model has
been developed for the distribution and composition of primordial gas
in a protogalactic disk that is in chemo-dynamical equilibrium, in the
presence of a photodestructive and photoheating UV---X-ray cosmic
background radiation (CBR) field.  The specific goal is to determine
where gravothermal instability (and hence SF) is initiated through the
\htwo\ cooling channel.  Additionally, where \htwo\ cooling cannot be
checked by the CBR alone, we place a lower bound on the global star
formation rate (SFR) in these objects by determining the additional
interstellar radiation field (ISRF) required to restore thermal
balance.

The discussion proceeds as follows.  First, an overview of the
argument is provided in \textsection\ref{ch:overview}, as well as
physical motivation for our model for a protogalaxy.  In
\textsection\ref{ch:sfcriterion}, the link between \htwo\ cooling and
Toomre instability is discussed, and we advance our specific criterion
for SF.  We provide a technical discussion of the processes by which
we determine the gas distribution and composition in
\textsection\ref{ch:distribution} and
\textsection\ref{ch:composition}, respectively.  The second
calculation, in which we estimate the global SFR, is then described in
\textsection\ref{ch:mstar}.  After a discussion of fiducial parameter
choices in \textsection\ref{ch:parameters}, the results of our
modelling procedure are given in \textsection\ref{ch:results}, and
discussed in \textsection\ref{ch:discussion}.  A summary of the
salient and novel aspects of this work can be found in
\textsection\ref{ch:summary}.

\section{Overview and \\ Supporting Calculations}
  \label{ch:overview}

\subsection{Overview of the Argument}
\label{ch:argument}

Our primary goal is to determine whether a putative dark galaxy is
gravothermally stable in its long time/steady state configuration.
Where this is the case, it is conceivable that such an object
(assuming that it can form) might remain dark indefinitely; otherwise,
from this point, if not well before, the galaxy will inevitably `light
up'.

Whereas rotational support stabilises against the collapse of
larger-scale density fluctuations, small-scale fluctuations can only
be stabilised by thermal pressure (see
\textsection\ref{ch:sfcriterion}).  Once the galaxy formation process
is complete, the level of rotational support becomes fixed; stability
is then principally determined by temperature, $T$.  In particular, as
will be shown in \textsection\ref{ch:sfcriterion}, most low-mass
protogalactic disks are stable at $T = 10^4$ K, while essentially all
will become unstable if they are able to cool to $T \sim 300$ K.

The final temperature of the gas depends on its composition.  Internal
cooling of the gas proceeds by radiative de-excitation of
collisionally excited atoms/molecules.  Whereas \HI\ cooling processes
rapidly become inefficient (in the sense that the timescale for
cooling becomes greater than the age of the universe) as collisional
excitation ceases below $\sim 10^4$ K, \htwo\ cooling remains
efficient above $\sim$ 300 K.  The key to initiating gravothermal
instability and SF in low mass disks is therefore efficient \htwo\
cooling; conversely, the prevention of SF requires the prevention of
\htwo\ cooling.

Within this paradigm, SF is thus a process that must be prevented,
rather than initiated: a source of dissociating radiation is required
to prevent the partial conversion of \HI\ to \htwo .  Within the
model, the CBR is the agent for dissociation, as well as the source of
heating that opposes \htwo\ cooling.  SF is deemed to be preventable
wherever the rate of photoheating due to the CBR exceeds the rate of
\htwo\ cooling within the gas; otherwise SF is inevitable.

We thus reduce the problem of determining the in/stability of the
protogalactic disk to determining the in/efficiency of \htwo\ cooling:
we assume that gravothermal stability implies and is implied by
thermal balance at $T \sim 10^4$ K.  (We will be able to explicitly
check the converse assumption --- that thermal instability at $T =
10^4$ K implies cooling to Toomre unstable temperatures --- in
\textsection\ref{ch:tmin}.)  Because we are interested primarily in
the situation of permanent gravothermal stability, we first assume
that the gaseous disk is in dynamical and chemical equilibrium at
$T=10^4$ K; \ie\ a long-time configuration at the end of \HI
-dominated thermal evolution.

This configuration represents the long-time state of a galaxy that has
managed, for whatever reason, to avoid SF, since the gas is Toomre
stable so long as \mbox{$T \gtrsim 10^4$ K;} if, in this
configuration, the gas cannot initiate \htwo\ cooling, this model
therefore represents the final state of a galaxy without stars.
Conversely, where \htwo\ cooling is inescapable, the result will be
thermal, chemical, and dynamical instabilities, leading necessarily to
gravothermal collapse and SF.

In other words, following authors like \citet{CorbelliSalpeter},
\citet{HaimanThoulLoeb}, \citet{OhHaiman}, and \citet{Schaye}, we link
SF to the existence of a cold phase within the gas, and examine under
what conditions the transition to the cold phase is initiated.  [See,
too, \citet{ParravannoElmegreen}, but also \citet{YoungEtAl}.]  Note
that turbulent effects, which can both initiate and prevent SF, are
neglected; all discussion of this point is deferred until
\textsection\ref{ch:turbulence}.  We have thus constructed a static
problem --- we make no attempt to track the evolution of the gas at
any stage.

Wherever possible, our assumptions have been tailored to minimise the
rates of both \htwo\ production and \htwo\ cooling, so as to make SF
as difficult as possible.  We thus place a firm lower limit on the
mass of gas that can exist in a steady state without undergoing any
SF.

Where we find that SF cannot be prevented by the action of the CBR, we
perform a second calculation to place a lower limit on the expected
SFR.  This is done by introducing a diffuse ISRF as a second source of
photoheating and photodissociation.  Specifically, we determine the
minimum ISRF intensity required to maintain thermal balance at $T =
10^4$ K, and so restore gravothermal stability.  With some simple
assumptions (see \textsection\ref{ch:mstar}), this ISRF can then be
related to a minimum global SFR within the model galaxy.

\subsection{The Back of the Envelope Calculation}
\label{ch:sbcrit}

The CBR can act to check \htwo\ cooling in two ways: by direct
photodestruction of \htwo\, and by overwhelming any \htwo\ cooling
with photoheating.  The cloud's only defence is self-shielding; \ie\
absorption of enough of the CBR by the outermost gas to shield inner
regions from its influence.

The action of the CBR is thus to prevent SF in a shielding `skin' that
covers both faces of a protogalactic disk, in a manner analogous to
photodissociation regions around UV sources in our own galaxy
\citep{HollenbachTielens}; SF is only possible in the self-shielded
centre.  The thickness of this skin can be characterised by the
critical surface density of gas, 
$\Sigma\subcrit$, below which the CBR is able to keep
the gas completely photodissociated.

We obtain a first estimate for $\Sigma\subcrit$ by noting that for the
CBR to prevent the production of \htwo , the total rate of
photodestruction in a column (which is just the flux of
photodestructive photons) must exceed the total recombination rate;
\viz:
\begin{equation}
\frac{J_{\nu}}{h\nu} ~ \Delta \nu \gtrsim \dot{n} ~ Z_d ~.
\label{eq:shieldcond}\end{equation}
Here, the flux of destructive photons is approximately the product of
the number of photons at the threshold frequency, $J_{\nu}/h\nu$
(where $J_{\nu}$ is the CBR flux at the threshold frequency $\nu$),
and the frequency interval over which photodestructive absorption can
occur, $\Delta \nu$.  Similarly, the recombination rate in the column
is approximately the product of the recombination rate per unit
volume, $\dot{n}$, and some characteristic column height, $Z_d$.

With the approximations $\Sigma \approx \rho_0 Z_d$ (where $\rho_0$
is, for now, a characteristic density), and $\dot{n} \approx
\alpha x_i x_j \rho_0^2 / \mu^2$ ($\alpha$ is the rate coefficient of
the reaction between species $i$ and $j$ that replenishes the cloud's
\htwo, $x_i$ is the relative abundance of species $i$, and $\mu
\approx 1.22 m_p$ is the mean molecular mass of particles in the gas),
it is possible to rewrite inequality (\ref{eq:shieldcond}) as:
\begin{equation}
\Sigma \lesssim \left(\frac{J_{\nu} ~ \Delta\nu}{h\nu}~
\frac{\mu^2}{\alpha x_i x_j} ~ \frac{\sigma^2}{2\pi G}\right)^{1/3}
= \Sigma\subcrit ~ .
\label{eq:critsb}\end{equation}
In writing this expression, we have also made use of equation
(\ref{eq:Z0}) to relate $\rho_0$ and $Z_d$, which introduces the gas
velocity dispersion, $\sigma$.


Using equation (\ref{eq:critsb}), we can now obtain a quantitative
estimate for $\Sigma\subcrit$, assuming $T = 10^4$ K (which fixes
$\sigma$ and the value of $\alpha$), chemical equilibrium (which
similarly fixes $x_i$ and $x_j$), and a putative CBR spectrum
$J(\nu)$.  To estimate $\Delta\nu$, we assume that \htwo\ shielding
occurs between the Lyman-Werner and Lyman-$\alpha$ threshold energies,
11.2---13.6 eV. 
With these numbers, we calculate $\Sigma\subcrit
\approx$ 3---4 M\sun\ pc$^{-2}$, depending on the character of the
CBR spectrum (see \textsection\ref{ch:cbrpars}).

In reality, the CBR will produce at least two layers of skin: one in
which it prevents \HI\ production (\ie\ an ionised layer) and,
internal to that, one in which \htwo\ production is prevented (\ie\ a
neutral, atomic layer).  Using the same argument as above, we find
that the surface density of the ionised layer is $\approx$ 1 M\sun\
pc$^{-2}$.

We therefore expect that the CBR will completely prevent SF where the
surface density of the protogalactic disk is less than several M\sun\
pc$^{-2}$.  [For comparison, observations are consistent with a fixed
SF threshold of $\sim 10$ M\sun\ pc$^{-2}$ (Taylor et al. 1994; see
also Martin \& Kennicutt 2001 and discussion in Schaye 2004).]

\subsection{Characteristic Timescales \label{ch:times}}

We assume chemo-dynamical equilibrium in our modelling.  Thus, while
our results do not rely on any particular galaxy formation scenario,
they do depend on these equilibrium states being accessible to each
system within its present lifetime.  Our model does not apply to {\em
forming} galaxies, nor does it attempt to follow the behaviour of the
gas as it transits from warm to cold.

In particular, our criterion for SF implicitly assumes that enough
time is available for the gas to cool from $T \sim 10^4$ K to $T \sim
300$ K after the onset of \htwo\ cooling.  In the remainder of this
section, we present some simple calculations in support of this
assumption.

First, we can evaluate the timescale for cooling \mbox{$t_c =
\varepsilon / \dot{\varepsilon}$}, where $\varepsilon \sim nkT$ is the
thermal energy density ($n$ is the number density of the gas; $k$ is
Boltzmann's constant) and $\dot{\varepsilon} = \Lambda$ is the net
cooling rate per unit volume.  Using the rates of cooling
listed in Appendix \ref{ch:rates}, and assuming gas composition in
chemical equilibrium at $T = 10^4$ K
(see Figure \ref{fig:eqcomps}), we find \mbox{$t_c \approx 1.5 \times
10^{-3} ~ n^{-1} ~ t_H = 20 ~ n^{-1}$ Myr} where \mbox{$t_H \approx
13.7$ Gyr} is the Hubble time and $n$ has units of cm$^{-3}$.  (Close
to the galactic plane, $n$ is typically in the range
$10^{-2}$---$10^1$ cm$^{-3}$; see Figure \ref{fig:egmodel}.)

Under these conditions, the \HI\ and \htwo\ cooling rates are
approximately equal.  In contrast, the same composition at 8000 K
leads to cooling times for \HI\ and \htwo\ cooling of $2.3 ~ n^{-1} ~
t_H$ and 15 $n^{-1}$ Myr, respectively: \htwo\ cooling remains
efficient well below the point at which \HI -driven thermal evolution
effectively ceases.

As we argue in \textsection 5.3, in gas near thermal equilibrium, the
\htwo\ abundance will always approach equilibrium from above; in this
case the equilibrium value therefore provides a lower limit to the
actual \htwo\ abundance.  For an \htwo\ abundance ten times greater
than the equilibrium value, $x\subhtwo \sim 10^{-3}$, the cooling time
at $T = 10^4$ K is reduced by a factor of 30 to $\sim 0.67 n^{-1}$
Myr.  Among tidal dwarf galaxies, the only observed population of
presently forming galaxies, the \htwo\ abundance is typically as high
as $x\subhtwo \sim 0.20$ \citep{BraineEtAl}; the cooling time in this
case is reduced to just $\sim 10^{-3} ~ n^{-1}$ Myr.

Next, we can make a rough estimate for the time needed to produce such
an overabundance, \mbox{$t\subhtwo = x\subhtwo / \dot{x}\subhtwo$}.
Again assuming equilibrium abundances at 10$^4$ K, and using the rate
for the dominant \htwo\ production process (Reaction 12 in Appendix
\ref{ch:rates}), \mbox{$\dot{x}\subhtwo \approx 2 \times 10^{-17} ~
n^2$} s$^{-1}$, we find $t\subhtwo \approx 1.6 ~ n^{-1}$ Myr to reach
$x\subhtwo = 10^{-3}$.  This value is of the same order of magnitude
as if we were to take a semi-empirical rate for \htwo\ production on
dust grains at $T \lesssim 10^3$ K (two-body \htwo\ production becomes
inefficient at these temperatures), $\dot{x}\subhtwo \sim 10^{-17} ~
n^2 ~ \mathrm{s}^{-1}$ \citep{HollenbachTielens}.

Thus, as equilibrium at 10$^4$ K is disturbed, we have $t\subhtwo <
t_c$, and $x\subhtwo$ increases above the equilibrium value; $t_c$ is
accordingly diminished.  At the same time, $t\subhtwo$ increases as
the temperature falls, until $t_c \ll t\subhtwo$.  At this point
two-body chemical evolution effectively ceases, leaving a `freeze-out'
abundance of \htwo : time-integrated studies of shock-heated (\ie\ $T >
10^4$ K, $x_{\mathrm{e}^-} \approx 1$) gas cooling catastrophically below
10$^4$ K \citep{SnK87, OhHaiman} find that $x\subhtwo$ asymptotes to a
`universal' freeze-out abundance of order $10^{-3}$ at $T \sim 4000$
K.  With the provision that this freeze-out \htwo\ abundance can
survive photodissociation (see \textsection\ref{ch:tmin}) from this
point on, the gas continues to cool until $t_c > t_{\mathrm H}$ at $T
\lesssim 300$ K.

Once the warm ($T \sim 10^4$ K) gas initiates \htwo\ cooling, it will
therefore transit to a cold ($T \sim 300$ K) state in $\lesssim 20 ~
n^{-1}$ Myr.  This should be compared to the characteristic timescale
for the first stage of SF, the free-fall time for a collapsing cloud,
\mbox{$t_f = (3\pi /16G \overline{\rho})^{1/2}$} \mbox{$\sim 52 ~
n^{-1/2}$ Myr}.  Where \htwo\ cooling occurs, the cloud will therefore
cool on a timescale {\em at least} a few times faster than the
collapse timescale; depending on the actual \htwo\ abundance, the
factor is more likely $\gtrsim 30$.

\section{The Model --- I \\
  The Criterion for Star Formation}
  \label{ch:sfcriterion}

\begin{figure*}[t]
\centering \includegraphics[scale=1.]{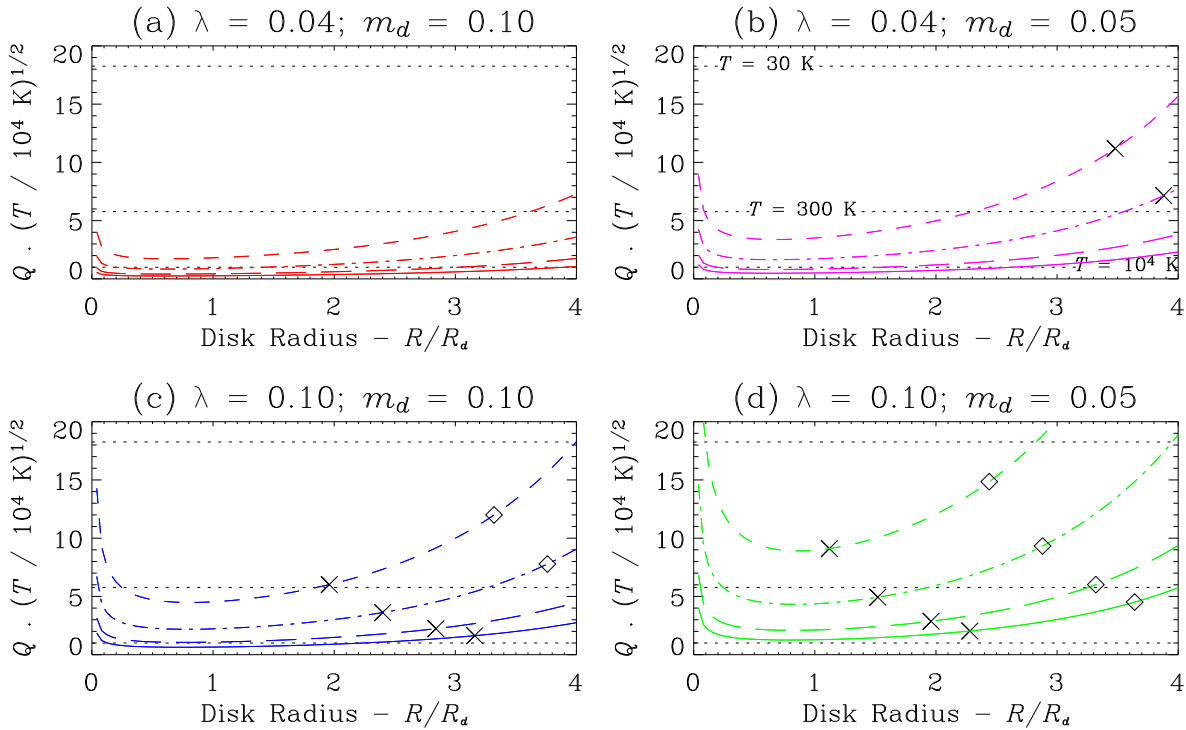}
\caption{\footnotesize{Toomre profiles $Q(R)$ for various parameter
  sets.  Panels (a) through (d) show the effect of different choices
  for the spin parameter, $\lambda$, and the disk mass fraction, $m_d$
  (see \textsection \ref{ch:parameters} for a detailed description);
  each panel differs in only one parameter from those adjacent to it.
  Within each panel, profiles are shown for virial masses of
  $1\times10^8$ M\sun\ (short-dashed line), $1\times10^9$ M\sun\
  (dot-dashed), $1\times10^{10}$ M\sun\ (long-dashed), and
  $5\times10^{10}$ M\sun\ (solid).  The Toomre parameter is
  temperature dependent: within each panel, dotted lines show where
  the Toomre threshold ($Q = 1$) would lie if the gas were at
  temperatures of $10^4$ K (bottom), 300 K (middle), and 30 K (top).
  Disks are unstable below the Toomre threshold.  To illustrate the
  relationship between Toomre stability and self-shielding, we also
  show the radius within which the gas is self-shielding (see
  \textsection\ref{ch:results}) for two different CBR spectral
  indices, $\eta = 0.7$ (diamonds) and 2.0 (crosses).  }
  \label{fig:toomre}}
\end{figure*}


The standard criterion for the gravothermal stability of a rotating
disk, the Toomre criterion \citep{Toomre}, can be obtained from the
dispersion relation for an axisymmetric density perturbation (\ie\ a
sound wave) with wavelength $\ell$ and frequency $\omega$
[\textsection 6.2.3 of \citet{BinneyTremaine}]:
\begin{equation}
\omega^2 = \kappa^2 - 2\pi G ~ \Sigma ~ \frac{2 \pi}{\ell} + 
\left(\frac{2 \pi}{\ell}\right)^2 c_s^2 ~ .
\label{eq:dispersion}\end{equation}
Here, $\kappa(R)$ is the epicyclic frequency of the orbit, and
$c_s(T)$ is the sound speed within the gas.  Since the general
waveform's time evolution is like $e^{-i\omega t}$,
stability requires that $\omega^2 > 0$, and hence:
\begin{equation}
\left(\frac{c_s\kappa}{\pi G \Sigma}\right)^2 \equiv Q^2 \ge 1 ~.
\label{eq:toomrecrit}\end{equation}
This is the Toomre stability criterion, written in terms of the Toomre
parameter, $Q$.

In order to determine the range of stable $\ell$s, we combine
equation (\ref{eq:dispersion}) with the condition $\omega^2 > 0$ to obtain:
\begin{equation}
\left(\frac{c_s\kappa}{\pi G \Sigma}\right)^2 >
4\left(\frac{\ell}{\ell\subcrit}
-\frac{\ell^2}{\ell\subcrit^2}\right) ~,
\label{eq:elltoomre}\end{equation}
where the left-hand side is now recognisable as $Q^2$.  Since $Q^2 \ge
0$, stability is assured for:
\begin{equation}
\ell > \ell\subcrit \equiv \frac{4\pi^2G\Sigma}{\kappa^2} ~ .
\label{eq:largeell}\end{equation}
Also, for $\ell^2 \ll \ell\subcrit^2$, inequality (\ref{eq:elltoomre})
reduces to:
\begin{equation}
\ell \lesssim \frac{c_s^2}{G\Sigma} ~,
\label{eq:smallell}\end{equation}
which is just the Jeans criterion for the gravothermal stability of a
thin disk (see also Schaye 2004).  These two cases can be interpreted
as the zero pressure and zero rotation limits, respectively.  Large
perturbations are stabilised by the effects of rotation (inequality
\ref{eq:largeell}), while small perturbations are stabilised by
internal pressure (inequality \ref{eq:smallell}).



Once the disk and halo have settled into dynamical equilibrium, all
gross dynamical quantities (\eg\ $\varrho_h$, $\Sigma$, $V$, $\kappa$)
become fixed.  The only variable quantity in inequality
(\ref{eq:toomrecrit}) that can influence $Q$ is then the sound speed,
$c_s = (kT/\mu)^{1/2}$.  The amount of the disk that is unstable
therefore depends strongly on $T$.  This is illustrated in Figure
\ref{fig:toomre}, which shows profiles of $Q(R)$ for the various
parameter sets that we will consider, and the regions in which these
disks would be unstable at temperatures of 10$^4$ K, 300 K, and 30 K.

As mentioned in \textsection\ref{ch:overview}, for low mass disks,
thermal stability at 10$^4$ K implies and is implied by gravothermal
stability; conversely, low mass disks only become Toomre unstable at
temperatures that are only accessible via the \htwo\ cooling channel.
Once \htwo\ cooling is initiated, the gas makes a rapid transition to
a cold phase with \mbox{$T \lesssim 300$ K} \citep{NormanSpaans}, in
which case it is virtually guaranteed (at least for $\mtwo \gtrsim
10^9$ M\sun) to be Toomre unstable.  For these reasons, we use the
\htwo\ cooling rate of the gas --- in chemical equilibrium and at $T =
10^4$ K --- in conjunction with the Toomre criterion for gravothermal
instability to define our criterion for SF.


Specifically, at a given point, star formation is deemed to occur
wherever the \htwo\ cooling rate exceeds the total rate of
photoheating, provided that the gas be Toomre unstable at $T \ge 300$
K.  Figure \ref{fig:toomre} illustrates the relation between these two
criteria: gravothermally unstable regions lie below the Toomre
threshold; we also show the radius within which the gas is
self-shielding.  

\section{The Model --- II \\
  The Distribution of Matter}
  \label{ch:distribution}

In the next three sections, we provide a detailed, technical
discussion of our three-step modelling procedure.  The means by which
we determine the distribution of matter in the protogalactic disk is
described in this section.  We discuss the solution for the
equilibrium chemical composition of the gas in
\textsection\ref{ch:composition}, and the calculation of the global
minimum SFR is discussed in \textsection\ref{ch:mstar}.  A reader
wishing to avoid such a technical discussion may choose to skip to
\textsection{\ref{ch:parameters}}.

The solution for the volume distribution of the gas involves two
separate calculations.  First, the analytic model of Mo, Mao \& White
(1998; hereafter MMW) is used to determine the radial distribution of
matter in the model protogalaxies.  The MMW model has been shown to
predict some global characteristics (\eg\ size distribution, velocity
profiles, Tully-Fisher relation) of nearby disk galaxies and damped
Lyman-$\alpha$ systems, based on more or less generic predictions of
$N$-body cosmological simulations (MMW).  Secondly, once the radial
mass distribution is known, the vertical distribution of mass in the
disk is determined using an extension of the formalism of
\citet{Bahcall} and \citet{DoveThronson}.  These two solution
processes are described in \textsection\ref{ch:scalelength} and
\textsection\ref{ch:scaleheight}, respectively.

In this section, and throughout the rest of this work, the symbols $R$
and $Z$ are used to represent the (cylindrical) galactocentric radius
and the vertical distance from the midplane, respectively; the
lowercase $r$ is reserved for spherical coordinates.


\subsection{The MMW Solution for the \\ Radial Mass Distribution
  of a Protogalactic Disk \label{ch:scalelength}}


The MMW model operates in the `adiabatic limit', in which the galaxy
formation process conserves an `adiabatic invariant'
\citep{BarnesWhite}.  The final configuration is thus made
independent of the assembly process, making it possible, for a known
initial mass distribution, to determine the final configuration,
provided the form of the final disk mass distribution is specified.

Following MMW, the dark and baryonic matter are assumed to follow the
same initial mass distribution; specifically, the fitting formula
proposed by Navarro, Frenk \& White (1997; hereafter NFW):
\begin{equation}
\rho_h(r) = \frac{\rho_{\mathrm{crit}} ~ \delta_c}{{r}/{r_h} ~ 
(1+{r}/{r_h})^2} ~ ,
\label{eq:nfw} \end{equation}
which has been shown to describe equilibrated halos in $\Lambda$CDM
cosmological simulations.  

The halo scale length, $r_h$, is defined in relation to the quantity
$r_{200}$, which is chosen such that the mean density within a sphere
centred on the halo and with radius $r_{200}$ is 200 times the
critical density for closure, \mbox{$\rho_{\mathrm{crit}}(z) =
3H^2(z)/8\pi G$}; \ie:
\begin{equation}
M_{200} \equiv M(r_{200}) 
= 200~\rho_{\mathrm{crit}} ~ \frac{4}{3}\pi~r_{200}^3 ~ .
\label{eq:m200} \end{equation}
Numerical experiments show $r_{200}$ to approximately demarcate
between regions in which material is still in-falling, and those in
which material tends towards equilibrated (radial) orbits
\citep{ColeLacey}; $r_{200}$ and $M_{200}$ are therefore referred to
as the virial radius and mass, respectively.  [See
\citet{BullockEtAl2001} for a more detailed discussion of the relation
between $r_{200}$/$M_{200}$ and the virial radius/mass.]  The precise
relation between $r_{200}$ and $r_h$ is set by the concentration
parameter, $c = r_{200} / r_h$.  The normalisation factor $\delta_c$
in equation (\ref{eq:nfw}) is defined as:
\begin{equation}
\delta_c \equiv \frac{200}{3} ~ c^3  \left(\ln(1+c) 
- \frac{c}{1+c}\right)^{-1} ~ .
\label{eq:delc} \end{equation}

Secondly, motivated by the observed distribution of {\em light} in
nearby galaxies \citep{Freeman}, the assembled {\em gaseous} disk is
assumed to have an exponential surface density profile, $\Sigma(R)$:
\begin{equation}
\Sigma(R) = \frac{m_d ~ M_{200}}{2 \pi ~ R_d^2} ~ e^{-{R}/{R_d}} ~ ,
\end{equation}
where the normalisation factor is chosen so that the total disk mass
is a parametric fraction, $m_d$, of the total dark-plus-baryonic mass.
Thus, for a given $M_{200}$ and $m_d$, $\Sigma(R)$ is completely
specified by a scale length, $R_d$.

The iterative MMW solution for $R_d$ relates the total energy of the
pre-assembly halo, $E$, to the angular momentum of the assembled disk,
$J_d$, using the  definition of the spin parameter, $\lambda$:
\begin{equation}
\lambda^2 \equiv \frac{J^2 ~ |E|}{G^2 ~ M_{200}^5} ~ ,
\label{spin} \end{equation}
where $J_d = j_d J$ is a parametric fraction, $j_d$, of the total
angular momentum of the pre-assembly, dark-plus-baryonic matter halo,
$J$.  The final result is the following expression for $R_d$ (MMW):
\begin{equation}
R_d = \left(2 ~ f_c\right)^{-0.5} ~ f_R ~ 
\frac{j_d ~ \lambda}{m_d} ~ r_{200} ~ ,
\label{eq:rd}\end{equation}
where $f_c(c)$ and $f_R(\lambda,c,m_d,j_d)$ are dimensionless factors
of order unity, dependent on the exact shape of the initial and final
mass distributions, respectively.  Wherever directly comparable, our
results are indistinguishable from those of MMW and of \citet{Schaye}.

\begin{figure*}[t] \begin{center}
\includegraphics[scale = 1]{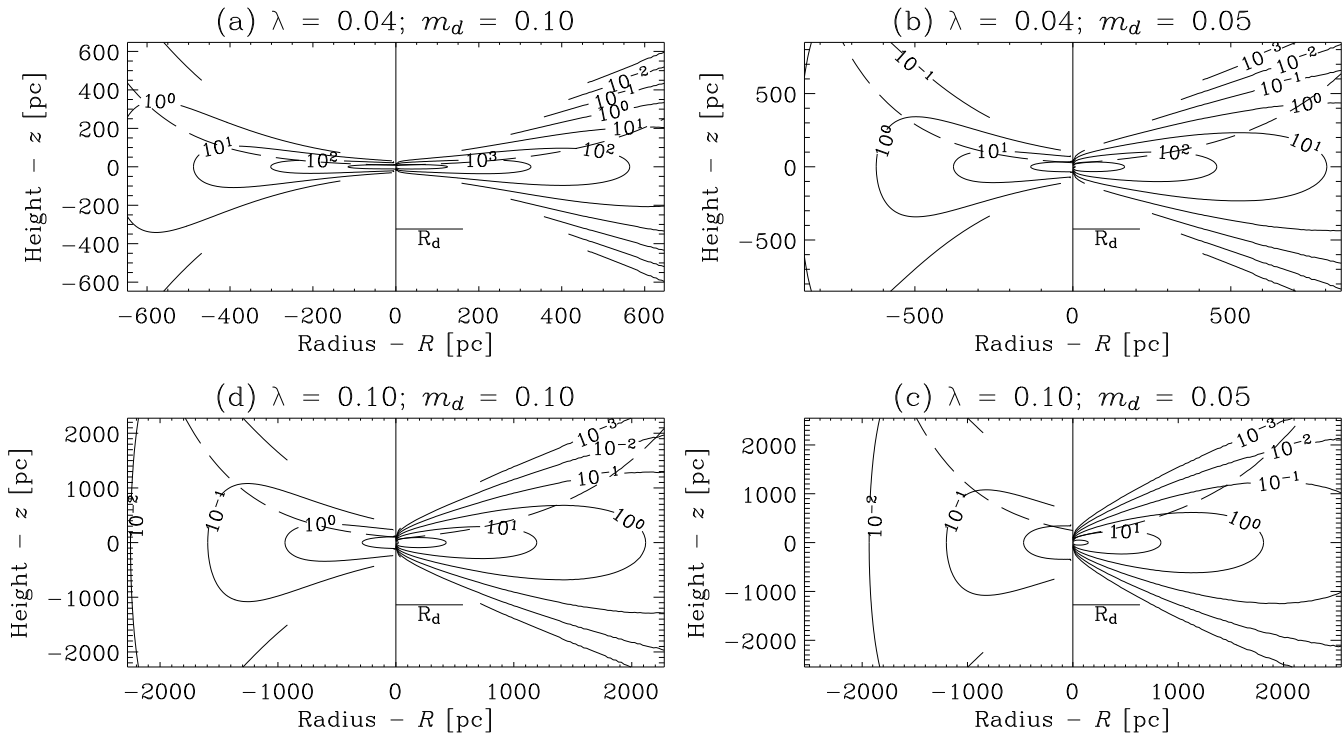} \end{center}
\caption{Mass distribution cross-sections for $M_{200} = 10^9$ M\sun ,
  illustrating the importance of the quantity $\xi$.  Contours are of
  constant density, in units of m$_\mathrm{p}$ cm$^{-3}$.  Each panel is
  8$R_d$ on a side.  The disk scale height $Z_d(R)$ is shown by the
  long-dashed line; profiles are computed out to 4$Z_d$.  Models where
  $\xi$ is computed as described in \textsection\ref{ch:scaleheight}
  are shown on the right of each frame.  For comparison, models with a
  \citet{Spitzer} sech$^2$ profile are shown on the left.}
\label{fig:xi} \end{figure*}


\subsection{The Vertical Mass Distribution \\
  of the Assembled Disk \label{ch:scaleheight}}

Once $\Sigma(R)$ is known, at a fixed $R$, the vertical density
profile for the disk gas, $\rho_d(R,Z)$ is governed by two equations
--- the Poisson equation:
\begin{equation}
\frac{1}{R}\frac{\partial}{\partial R} \left[R~\nabla
\Phi_R\right] + \frac{\partial^2 \Phi}{\partial Z^2} 
=  4 \pi G ~ (\varrho_h + \rho_d) ~ ,
\label{eq:poi} \end{equation}
and the equation of hydrostatic equilibrium (in the $Z$ direction):
\begin{equation}
\sigma^2 ~ \frac{\partial \rho_d}{\partial Z} 
+\rho_d ~ \frac{\partial \Phi}{\partial Z} = 0
%
\label{eq:boltz}\end{equation}
In these equations, $\Phi(R,Z)$ is the gravitational potential,
$\Phi_R(R,Z)$ is the radial component of the potential, $\varrho_h(r)$
is the density of the dark matter halo (determined numerically as in
Appendix B), and $\sigma$ is the (thermal) gas velocity dispersion.
Since the gas is assumed to be isothermal, $\sigma$ is taken to be
constant throughout the disk.

Using the equation of hydrostatic equilibrium is equivalent to using
the first moment in $\overline{v_Z}$ of the collisionless Boltzmann
equation, with the velocity dispersion tensor, $\sigma_{ij}^2$ (see,
\eg, \textsection4 of Binney \& Tremaine, 1987), set to
$\delta_{ij}\sigma^2$.  In the context of non-interacting stellar
dynamics (\cf\ an equilibrated gas), this assumption would introduce
an error of order $(Z\,Z_d/R\, R_d)$ \citep{Bahcall}.

The disk is assumed to be in centrifugal balance in the $R$ direction,
and particles to travel stable, circular orbits; \viz\ $\nabla\Phi_R =
V^2 / R$, where $V(R)$ is the orbital velocity of the gas at a radius
$R$.  With this substitution in equation (\ref{eq:poi}), these two
equations can be combined by differentiating equation (\ref{eq:boltz})
with respect to $Z$, to give \citep{Bahcall, DoveThronson}:
\begin{eqnarray}
%
\left(\frac{1}{\rho_d} ~ \frac{\partial \rho_d}{\partial Z}\right)^2  - &&
\frac{1}{\rho_d}~\frac{\partial^2 \rho_d}{\partial Z^2} \nonumber \\ 
= \frac{4 \pi G}{\sigma^2} ~ \rho_d\ + &&
 \frac{4 \pi G}{\sigma^2} ~ \left(\varrho_h - \frac{1}{4 \pi G ~ R}
~ \frac{\partial V^2}{\partial R}\right) ~ .
\label{eqn:ode} 
\end{eqnarray}

Following \citet{Bahcall} and \citet{DoveThronson}, equation
(\ref{eqn:ode}) is made dimensionless by defining a scale height for
the disk $Z_d(R)$:
\begin{equation}
Z_d(R) \equiv 
\left(\frac{\sigma^2}{2\pi G~\rho_0(R)}\right)^{{1}/{2}}~, 
\label{eq:Z0} \end{equation}
where $\rho_0(R) = \rho_d(R,~0)$ is the volume density of the disk gas
on the plane, at a distance $R$ from the galactic centre.  [Note that
$Z_d$ is twice the exponential scale height \citep{Bahcall}.]  This
leads (again, for a fixed $R$) to an ordinary differential equation
for the dimensionless density \mbox{$y(x) = \rho_d(R,Z) / \rho_0(R)$},
where $x=Z/Z_d$:
\begin{equation}
\frac{d^2y}{dx^2} = \frac{1}{y} \left(\frac{dy}{dx}\right)^2
- 2 ~ y^2 - 2 ~ \xi(R, Z) ~ y ~ ,
\label{eq:yofx}\end{equation}
subject to the boundary conditions:
\begin{equation}
y(x = 0) = 1 ~~~~~~ \mathrm{ and } ~~~~~~ 
\frac{dy}{dx}\Big\vert_{x=0} = 0 ~.
\end{equation}

It is worth making a few comments here about the quantity $\xi$ in equation (\ref{eq:yofx}), which is defined as: 
\begin{equation}
\xi(R,Z) \equiv \frac{1}{4\pi G \rho_0} \left(4\pi G \varrho_h(r) -
\frac{1}{R} \frac{\partial V^2(R)}{\partial R} \right)~,
\label{eq:xi} \end{equation}
but which can be rewritten in a more convenient form (Appendix \ref{ch:xicalc}).
If $\xi$ is neglected altogether, then equation (\ref{eq:yofx}) is
analytic, yielding \mbox{$y(x) = \mathrm{sech}^2~x$} \citep{Spitzer}.
This formalism was first derived in the context of stellar
dynamics close to the galactic plane and at modest distances from the
galactic centre, in which case both $V$ and $\varrho_h$ vary slowly
across the volume of interest.  Authors therefore typically neglect
the $\partial V^2/\partial R$ term and treat $\varrho_h$ as constant
(\eg\ Maloney 1993; Dove \& Shull 1994).

This work, however, considers lower mass galaxies with manifestly
non-flat rotation curves.  Moreover, we are interested primarily in
the outermost gas, since this is where the effects of the CBR are
greatest; this is also where the effect of a non-zero $\xi$ is most
pronounced.  It is therefore inappropriate to adopt these standard
approximations here.  The effect of a non-zero $\xi$ is illustrated in
Figure \ref{fig:xi}, which shows several model gas distributions as
well as the equivalent distribution assuming $y = \mathrm{sech}^2 ~ x$.

In summary, the modelling procedure is as follows: $y(x)$ is
determined by numerical integration of equation (\ref{eq:yofx});
$\rho_0$ is then fixed with reference to the definition of
$\Sigma(R)$, \viz:
\begin{equation}
\rho_0 = \frac{2 \pi G ~ \Sigma^2(R)}
{\sigma^2 ~ \left( 2 \int_0^{\infty} dx ~ y(R,x) \right)^2} ~ .
\label{eq:rhod0} \end{equation}
The presence of $\rho_0$ in the definition of $\xi$, as well as the
implicit dependence of $\xi$ on $Z$, make it necessary to solve for
$y(x)$ iteratively: a trial value of $\rho_0$ is used to determine
the value of $\xi$ in the integration of equation (\ref{eq:yofx}); the
value of the integral is then used in equation (\ref{eq:rhod0}) to
obtain the next $\rho_0$.  This process rarely requires more than
several iterations for convergence.

\section{The Model --- III \\
  The Chemical Composition \\ of the Gas and the CBR}
  \label{ch:composition}

Once the distribution of the disk gas is known, we determine its
equilibrium chemical structure.  Nine chemical species are identified
within the gas, interacting via 21 collisional and nine
photodestructive processes, as outlined in \textsection\ref{ch:chem}.
The gas is subject to a UV---X-ray CBR field; we describe the CBR
parameterisation, including gas self-shielding, in
\textsection\ref{ch:cbr}.  The solution for the equilibrium chemical
composition of the gas is discussed in \textsection\ref{ch:comp}.

\subsection{Chemical Species and Processes
  \label{ch:chem}}

Following Haiman, Thoul \& Loeb (1996; hereafter HTL) and
\citet{AbelEtAl}, we distinguish nine chemical species within the gas: \hz, \hm, \hp,
\htz, \htp, \hez, \hep, \hepp, and \eminus , assuming a (primordial)
helium mass fraction of 0.24.  These chemical species are allowed to
interact via the 21 collisional processes identified by HTL.  In
addition, the nine photoionisation and photodissociation processes
listed in \citet{AbelEtAl} are included.  For ionisation of \hez\ and
\htz, the more recent cross-sections given by
\citet{YanSadeghpourDalgarno} are employed.  These expressions give
much lower cross-sections at high energies than those used by
\citet{AbelEtAl}, which reduces the rate of heating in marginally
shielded areas, but increases the penetration depth.  The rate
coefficients and cross-sections for all of these processes are listed
in full in Appendix \ref{ch:rates}.

We completely neglect the effects of any dust or metals present in the
gas, even though the IGM is expected to have seen considerable
enrichment \citep{AguirreEtAl, Schaye2000}.  Metals can make a
significant contribution to the cooling rate within the self-shielded
region \citep{GalliPalla, KatzWeinbergHernquist}; moreover, both dust
and metals act as sites for \htwo\ production.  These effects only
become dominant, however, once two-body collisional production of
\htwo\ becomes inefficient, well below the assumed temperature of
$10^4$ K \citep{BuchZhang, GalliPalla}.  Further, both dust and metals
can absorb UV CBR radiation, so contributing to the level of
self-shielding within the cloud.  The neglect of dust and metals
therefore minimises the predicted level of \htwo\ production and
cooling.

For each individual chemical species, an expression can be written for
the net rate of production/consumption in terms of the number
densities of all nine species. 
It is most convenient and instructive to cast these expressions in
terms of the relative chemical abundances, $x_i = n_i / n\subh$, where
$n_i$ is the number density of species $i$, and $n\subh$ is the
number density of hydrogen nuclei, \viz:
\begin{equation} 
\frac{dx_i}{n\subh ~ dt} 
= \sum_{p}{\bigg [}
\sum_{q}{\bigg (}
b_{pq}^{(i)}~\alpha_{p, q}~x_p~x_q {\bigg )} 
 + ~
d_{p}^{(i)}~\frac{\Gamma_{p}}{n\subh}~x_p ~
{\bigg ]} ~ .
\label{eq:dxidt} %
\end{equation}
Here, $\alpha_{p,q}$ is the temperature dependent rate coefficient for
the reaction between $p$ and $q$, $\Gamma_{p}$ is the rate of
photoionisation/photodissociation of $p$, and $b_{pq}^{(i)}$,
$d_{p}^{(i)}$ = 0, $\pm$1, or $\pm$2, depending on the number of $i$
particles produced/consumed in the reaction.  Note that in the high
density/zero radiation limit, the equilibrium abundances are
independent of density; they are determined by the temperature
dependent $\alpha$s alone.  Equation (\ref{eq:dxidt})
encapsulates a particularly `stiff' set of nine highly nonlinear
coupled differential equations for the chemical composition of the gas
at a fixed point (HTL).

\subsection{The Photodestructive Effects of the CBR
  \label{ch:cbr}}

We consider the CBR in the range \mbox{13.6 eV $\le h\nu \le$ 40 keV}
(\ie\ 912 \AA\ $\ge \lambda \ge 0.31$ \AA --- we find that
photoionisation of He and He$^+$ by soft X-rays is important in \htwo\
production, as well as being a significant photoheating mechanism.)
Within this range, the CBR spectrum, $J_{\mathrm{CBR}}(\nu)$, is
described using a simple power law, characterised by its spectral
index, $\eta$, and a normalisation constant, $J_{21}$, chosen such
that:
\begin{eqnarray}
J_{\mathrm{CBR}}(\nu,\, R,\, Z) = J_{21} ~ 
\left(\frac{\nu}{\nu\subHI}\right)^{-\eta} ~ e^{-\tau(\nu; R, Z)} \nonumber \\ \times 10^{-21} \mathrm{~erg^{-1}~s^{-1}~Hz^{-1}~cm^{-2}~Sr^{-1}} ~ .
\label{eq:cbr} \end{eqnarray}
Here, $\nu\subHI$ is the frequency of the Lyman-$\alpha$ photon:
$h\nu\subHI \approx 13.6$ eV.  This parameterisation is the simplest
possible; to assess its relevance, we direct the reader to the
compendium of CBR observations across the whole electromagnetic
spectrum provided by \citet{Henry}.  We will trial two extreme values
of $\eta$ (see \textsection\ref{ch:cbrpars}), in an attempt to bracket the real situation.

The frequency dependent optical depth, $\tau$, defined as:
\begin{equation}
\tau(\nu;\,R,\,Z) = \sum_{i}\bigg [a_{i}(\nu) 
\int^{\infty}_{Z}dZ' ~ n_{i}(R,Z') \bigg ] ~ ,
\label{eq:tau} \end{equation}
is used to modulate the CBR spectrum as it penetrates into the cloud,
so mimicking radiative transfer.  Here, $a_i(\nu)$ is the frequency
dependent cross-section for photodestruction of species $i$, as given
in Appendix \ref{ch:rates}.  Note that this treatment of radiative
transfer is `one way' --- no attempt has been made to account for
radiation that crosses the midplane --- leading to a very minor
underestimation of the photoheating rate close to the plane.

We have made three substantial omissions in writing equations
(\ref{eq:cbr}) and (\ref{eq:tau}).  First, any extinction of the CBR
due to the presence of dust or metals has been neglected.  Secondly,
secondary ionisations by energetic electrons from X-ray ionisations
are ignored.  While these ionisations have no effect on the heating
rate, the enhanced electron abundance would increase the \htwo\
production rate (and the total rate of cooling); the omission of
secondary ionisations will therefore again lead to a lower limit on
the shielded fraction.  Thirdly, and most importantly, diffuse
radiation produced by the gas itself has been ignored; that is, any
photons (re-)radiated by the gas are assumed to be free to escape
unhindered.  This assumption is justified for the optically-thin
outermost gas; its validity within the self-shielded region is
certainly problematic, but a detailed treatment of radiative diffusion
through the cloud would be dependent on both the thermal and dynamic
evolution of the disk during its formation.  As has been stated, this
is beyond the scope of the present undertaking.

For each photoreaction except Solomon photodissociation of \htwo\ (see
below), the photodissociation/photoionisation rate, $\Gamma_i$, is
then given by:
\begin{equation}
\Gamma_i = 2 \pi \int_{\nu_i}^{\infty} d\nu ~
\frac{J(\nu)}{h\nu} ~ a_i(\nu) ~ ,
\label{eq:gamma}\end{equation}
where $\nu_i$ is the threshold energy for dissociation of species $i$,
and the factor of 2$\pi$ arises from integration over all angles of
incidence onto the flat disk.

Photodissociation of \htwo\ occurs primarily in the Lyman-Werner bands
(11.26---13.6 eV) via the two-step Solomon process
\citep{StecherWilliams}:
\begin{equation}
\mathrm{H}^0\ + h\nu \rightarrow\ \mathrm{H}_2^* \rightarrow\ 2\mathrm{H}^0 ~.
\end{equation}
This process is harder to deal with than other photoprocesses, since
the structure of the molecular energy levels (and so the cross-section
for photodestruction) is so much more complex.  Moreover, Doppler
broadening causes individual lines to overlap.  Following
\citet{AbelEtAl}, absorption is therefore restricted to the Lyman band
(12.24---13.51 eV); the CBR spectrum is treated as flat in this narrow
range.  Thus, rather than explicitly considering \htwo\ dissociating
photons longward of 912 \AA, the \htwo\ photodissociation rate is
determined with reference to the mean number of photons near 1000 \AA.
The penetration probability formalism of \citet{deJongDalgarnoBoland},
which modulates the unextinguished CBR field with a shielding factor,
$\beta$, is used to simulate \htwo\ self-shielding,
including the effects of Doppler broadening; the adopted expression
for $\beta(\tau\subhtwo)$ is given in Appendix \ref{ch:rates}.

\subsection{Chemical Composition in Equilibrium
  \label{ch:comp}}

\begin{figure}[b] \begin{center}
\includegraphics[width=8cm]{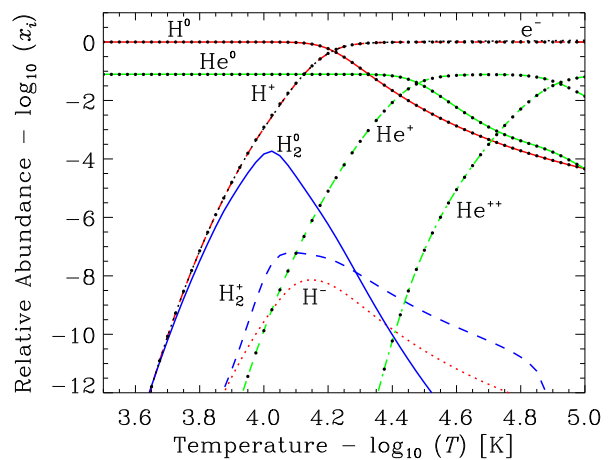} \end{center} 
\caption{\footnotesize{Equilibrium abundances of nine chemical species
    in the absence of any photodestruction.  Chemical species are as
    labelled.  The points show the equilibrium solutions for the
    reduced set of equations in which \hm, \htz, and \htp\ are
    omitted.  All abundances are scaled relative to the total hydrogen
    nucleus number density, $n\subh$.  Note that the electron
    abundance closely tracks the proton (\hp ) abundance.  This plot
    is provided only as a diagnostic for our solution procedure, and
    should not be mistaken for actual evolutionary tracks.
    (see \textsection\ref{ch:comp}). \label{fig:eqcomps}}}
\end{figure}

Once the radiation density at a given point is known (and hence the
rate coefficients for the various photochemical processes), it is
possible to find the equilibrium chemical composition at that point by
solving the system of equations represented by equation
(\ref{eq:dxidt}), with the time derivatives on the left-hand side set
to zero.  We do this using a modified Netwon-Raphson algorithm
(\textsection 9 of Press et al. 1992), where we compute the Jacobi
matrix analytically.  For $dx_i/dt = 0$, only six of the nine
equations remain independent; it is also necessary to ensure that the
total number of hydrogen nuclei, helium nuclei, and electrons are
conserved separately \citep{KatzWeinbergHernquist}.

As a diagnostic for the solution procedure {\em only}, Figure
\ref{fig:eqcomps} shows the equilibrium chemical composition (in the
absence of photodestruction) as a function of temperature.  In this
plot, the lines show our steady-state solutions for the full set of
equations encapsulated by equation (\ref{eq:dxidt}); for comparison,
the points show algebraic solutions for a reduced system of six
equations obtained by omitting \hm, \htz, and \htp.  Note that these
results differ significantly from those in the directly comparable
Figure 2 of HTL, which were obtained by direct integration of equation
(\ref{eq:dxidt}) until apparent convergence.  In particular, our
predicted \htz\ abundance is one to two orders of magnitude lower than
that of HTL, depending on the reaction rate expressions used (see
Appendix \ref{ch:rates}).  The disagreement arises from the fact that
the results shown in HTL had not yet fully converged (Zolt\'an Haiman,
private communication --- note that this does not affect the
conclusions of HTL), as can be readily verified by direct
substitution.

It should also be noted that, contrary to \citet{HaimanReesLoeb},
assuming chemical equilibrium is expected to lead to a lower limit on
the actual \htwo\ abundance at $T=10^4$ K, and by extension on the
\htwo\ cooling rate.  The assumed rate coefficient for \htwo\
production is temperature independent, while that for consumption
decreases with temperature; thus, the effect of a small decrease in
temperature is an increase in the {\em net} rate of \htwo\ production
[see, \eg, Figure 4 of \citet{AbelEtAl}].  The net rate of \htwo\
production, therefore, can only go to zero if the \htwo\ abundance is
dropping (or if the gas temperature is rising).  The \htwo\ abundance
in an equilibrating gas will therefore always reach equilibrium
\textit{from above}.  (For a time-evolved demonstration of this point,
see Figure 4 of Haiman, Abel and Rees 2000.)

Moreover, the situation of chemical equilibrium for \htwo\ is
physically improbable for temperatures between $\sim 10^3$---$10^4$ K,
because in the absence of a source of photoheating, the gas will be
thermally unstable; gas is far more likely to cool past these
temperatures without equilibrating.  In this sense, we emphasise the
limited physical relevance of Figure \ref{fig:eqcomps}, which is
provided only as a diagnostic, and refer the reader to plots of
nonequilibrium abundances in cooling and collapsing gas clouds like
those found in Figures 3---5 of \citet{OhHaiman}, Figure 3 of
\citet{AbelEtAl}, or Figure 6 of \citet{SnK87}.

\begin{figure*}[t]
\centering \includegraphics[scale = 1.]{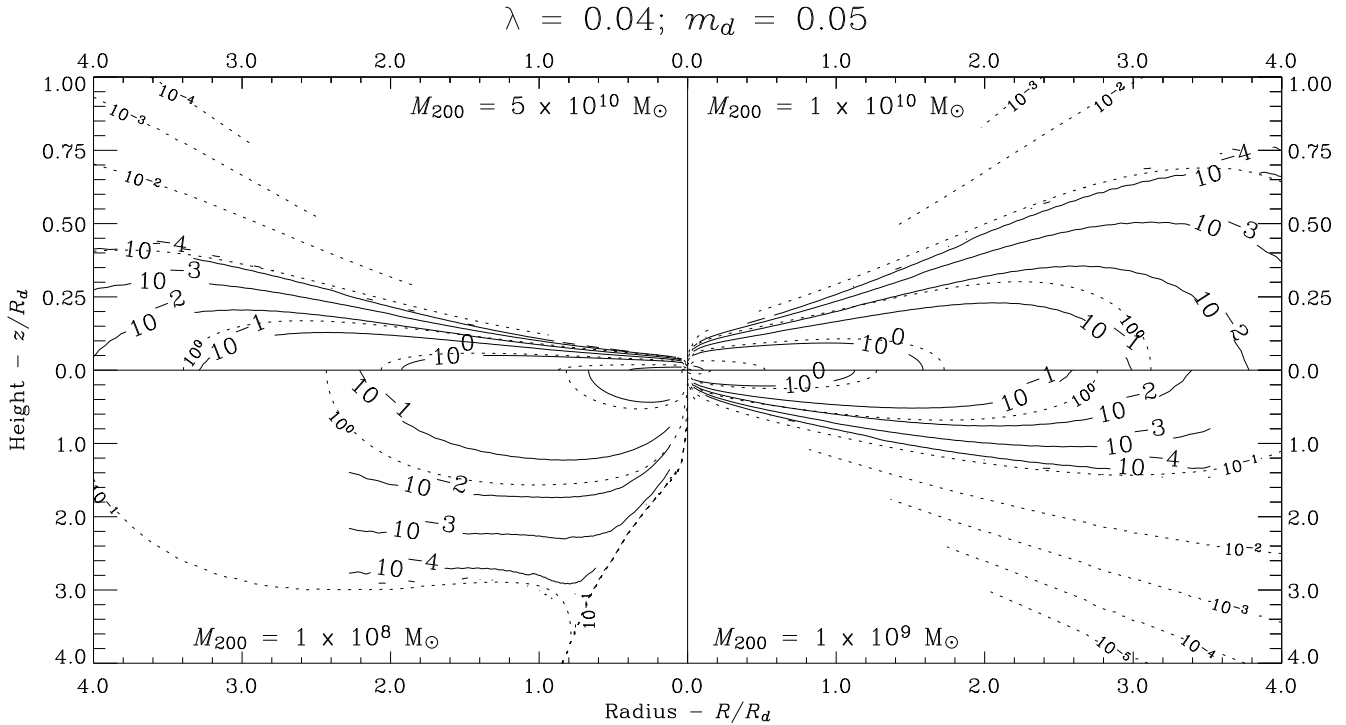}
\caption{\footnotesize{Contours of $u_d$, shown in cross section.
    $u_d$ is the minimum ISRF energy density (at 1000 \AA) required
    for thermal balance, determined as per
    \textsection\ref{ch:mstar}. Contours are shown (in units of the
    Habing flux) for example model ($\lambda,~m_d$) = (0.04, 0.05);
    clockwise from top-left are models with $\mtwo =
    5\times10^{10},~10^{10}, ~10^{9}$, and $10^{8}$ M\sun.  Note the
    different ranges of the vertical axes in the upper ($R_d$) and
    lower ($4R_d$) panels.  We only solve for $u_d$ at radii where the
    gas is Toomre unstable at $T \ge 300$ K (see Figure
    \ref{fig:toomre}).  Overlaid are contours of $n\subh$ (dotted
    lines), in units of cm$^{-3}$.
    \label{fig:egmodel}}}
\end{figure*}

\section{The Model --- IV \\
  The Minimum SFR}
  \label{ch:mstar}

At each point where \htwo\ cooling is found to be sufficient to
initiate gravothermal instability, we introduce an additional internal
radiation field and determine the minimum ISRF required to preserve
thermal balance at $10^4$ K; it is then possible to relate this ISRF
to a minimum global SFR, $\Rmin$.  In other words, we solve for the
minimum ISRF-regulated SFR within the cloud.

This is done by including a tenth equation in the system of equations
described in \textsection\ref{ch:comp}, requiring that the collisional
\htwo\ cooling rate ($\Lambda\subhtwo$, given in Appendix
\ref{ch:rates}) be balanced by the net rate of photoheating, \viz:
\begin{equation}
\sum_i \zeta_i - \Lambda\subhtwo = 0 ~.
\end{equation}
Here, $\zeta_i$ is the rate of photoheating due to absorption by
species $i$, which (for all processes except Solomon dissociation of
\htwo --- see Appendix A) is given by:
\begin{equation}
\zeta_i = n_i ~ \int_{\nu_i}^{\infty} d\nu ~ (h\nu - h\nu_i) ~
\frac{J(\nu)}{h\nu} ~ a_i(\nu) ~ ,
\end{equation}
where the radiation flux $J = J_{\mathrm{CBR}} + cu_d$ now has two
components.  The first is due to the CBR, where $J_{\mathrm{CBR}}$ is
defined by equation (\ref{eq:cbr}); the second is the new diffuse
ISRF, characterised by the diffuse radiation energy density, $u_d$.
Note that we use this same $J$ in equation (\ref{eq:gamma}), so that
the ISRF maintains thermal balance by some combination of photoheating
and \htwo\ photodissociation.

The spectral shape of $u_d$ is fixed with reference to the stellar
population modelling utility Starburst99
\citep{Starburst99}\footnote{Specifically, we used the data shown in
Figure 2b, available at http://www.stsci.edu/science/starburst99/.},
which gives the SED for continuous star-formation, $L_{99}(\nu)$, with
units of erg s$^{-1}$ Hz$^{-1}$ \mbox{(1 M\sun\ yr$^{-1})^{-1}$}.  We
adopt a standard SF scenario, which assumes a Salpeter IMF over a mass
range of \mbox{1---100 M\sun}, and a metallicity of 0.020 (Z\sun); our
results are not strongly dependent on the choice of metallicity.  Over
the spectral range of interest, this SED converges after $\lesssim 10$
Myr (a few times the lifetime of O---B stars); we use this converged
SED.  It should be noted that the Starburst99 SED only extends to 100
\AA ; since self-shielding against the diffuse ISRF is ignored, this
will lead to only a minor underestimate of the rate of \HeI\
ionisation, and so the rate of heating.

To convert this luminosity to a radiation density, it is necessary to
make an estimate of the mean lifetime of each photon,
$\tilde{t}_{\nu}$, which is done with reference to the mean free path
of a photon with energy $h\nu$, $\ell_{\nu} \equiv (n a_{\nu})^{-1} =
\tilde{t}_{\nu} c$.  With these substitutions, $u_d$ can be written:
\begin{equation}
u_d(\nu) = {\cal U} ~ L_{99}(\nu) ~ \left(c \sum_{i} n_i a_i(\nu)
\right)^{-1} ~ .
\end{equation}
While the spectral shape of $u_d$ is given by $L_{99}$, its intensity
is thus characterised by a scale factor, ${\cal U}$, which has the
units of a SFR per unit volume.

${\cal U}$ is thus taken as the tenth unknown in a system of ten
equations, representing the minimum SFR (per unit volume, and assuming
chemical equilibrium) required to produce a sufficient ISRF to
maintain thermal balance at $T = 10^4$ K, and so restore gravothermal
balance.  Note that ${\cal U}$ gives the SFR required to ensure the
stability of the volume $dV$, not the SFR within $dV$.  Figure
\ref{fig:egmodel} shows the resultant ISRF field in cross-section for
several exemplary models.  A global SFR, $\Rmin$, is then found by
integrating ${\cal U} = d\Rmin/dV$ over the whole disk.

Clearly, this is not a solution for the exact problem.  That said,
$\Rmin$ is expected to provide a reasonable estimate for the minimum
SFR in the cloud, since `maximum damage' has been allowed; photons
are delivered directly to wherever they are required.  Under these
conditions, so long as the actual SFR exceeds $\Rmin$, the
non-star-forming remainder of the disk will remain stable; otherwise,
some part of the disk will become unstable and the SFR will increase
as a result.


\section{The Model --- V \\
  Choice of Parameters}
  \label{ch:parameters}

The model now outlined consists of three components described by nine
parameters.  In this section, the fiducial values adopted for these
parameters are discussed.

Cosmology enters our calculations via $\rho_{\mathrm{crit}}$, which is
used to determine $r_{200}$ for a given $M_{200}$ (see equation
\ref{eq:m200}); we adopt the now standard WMAP cosmology
\citep{BennettEtAl, SpergelEtAl}.  Further, model disks are assumed to
form at zero redshift.  Since $\rho_{\mathrm{crit}}$ decreases with
decreasing redshift, $r_{200}$, and hence $R_d$, will be greater for
smaller disk formation redshifts (see equation 9); at a fixed $R /
R_d$, this minimises $\rho_d$ and so self-shielding and \htwo\
production, in keeping with the ethos of the model.

\subsection{The Progenitor Halo: $M_{200}$, $c$, $\lambda$}

The maximum virial mass considered is $\mtwo = 5 \times 10^{11}$
M\sun, typical of a nearby LSB galaxy.  The minimum mass, $\mtwo
\approx 10^8$ M\sun, is set by the form of the analytic model: below
this limit, the solution process for $Z_d$ and $\rho_0$ fails to
converge at $R \sim R_d$.  Mathematically, the long, slow rise of the
rotation curve for lower mass galaxies makes $\xi$ a large negative
number at these radii, which eventually drives the vertical density
profile, $y(x)$, to increase with distance off the plane (see equation
\ref{eq:xi}).  In this unphysical case, we truncate the integration,
and treat the stationary point as a sharp boundary.  It may be that
lower mass halos do not form exponential disks.

The concentration parameter, $c$, is chosen using a simple scaling
relation drawn from Figure 6 of \citet{NFW96}:
\begin{equation}
c = 7.5 \left(\frac{M_{200}}{M_*}\right)^{-1/9} ~ ,
\end{equation}
where $M_*$ is the (cosmology dependent) `nonlinear' mass, as defined
in that work.

Numerical experiments show that the spin distribution in hierarchical
clustering scenarios is well described by a `log-normal' distribution
centred at $\lambda_0 = 0.04$, and with width $\delta \lambda = 0.5$
\citep{WarrenEtAl, BullockEtAl}.  Two values for the spin parameter,
$\lambda$, are adopted to sample this distribution: a `typical' value
of 0.04, and a `high' value of 0.10.  These values approximately
correspond to the 50 and 90 percent points of the expected
distribution (see also MMW).

\subsection{The Disk: $m_d$, $j_d$, $\sigma$, $T$
  \label{ch:diskpars}}

Based solely on the WMAP cosmology \citep{SpergelEtAl}, baryons are
expected to make up 16 percent (by mass) of all matter.  However, not
all of the baryonic material initially within an overdense region will
ultimately settle into the disk.  Moreover, of the mass that does
collect at the centre of the potential well, some fraction may be
lost, \eg\ through evaporation after reionisation \citep{ShavivDekel}.
Two fiducial values for the disk mass fraction are therefore adopted,
$m_d = 0.05$ and 0.10 (see also MMW).

Assuming that the halo's angular momentum originates from tidal
torquing, it is natural to assume that both the dark and baryonic
components obtain the same specific angular momentum
\citep{FallEfstathiou}; thus, at the time of virialisation, $m_d =
j_d$.  While $N$-body and hydrodynamic simulations predict that
$j_d/m_d$ is significantly less than unity for assembled disks,
\citep{NavarroWhite, NavarroSteinmetz}, the MMW model requires that
$j_d/m_d$ be close to unity in order to produce disk sizes and
properties that are in agreement with observations.  Accordingly, we
assume $j_d = m_d$ for the assembled disk.

The velocity dispersion, $\sigma$, is assumed to be constant, with the
value of $8.2$ km s$^{-1}$; this corresponds to the thermal velocity
dispersion at $T = 10^4$ K.  Our chosen value is larger than the
canonical value of 6 km s$^{-1}$, which, using a recalibrated Toomre
criterion ($Q < 2$), has been shown to predict the location of star
formation thresholds in around 2/3 of nearby spiral galaxies
\citep{Kennicutt89, MartinKennicutt}.  However, \citet{Schaye} has
recently suggested that this agreement is largely coincidental: for
typical HSB disk galaxies only, the point where the disk is Toomre
unstable assuming \mbox{$c_s = \sigma = 6$ km s$^{-1}$} fortuitously
coincides with where the transition to the cold phase becomes
possible.

\subsection{The CBR Field: $J_{21}$, $\eta$}
\label{ch:cbrpars}

The CBR parameters are fixed using the value of the low-redshift \HI\
photoionisation rate, $\Gamma\subHI$, obtained by \citet{ScottEtAl}.
They give a best fit value of \mbox{$\Gamma\subHI = 2.0^{+9.0}_{-1.4}
\times 10^{-13}$ s$^{-1}$} for $z < 1$, where $\Gamma\subHI$ is
related to the UV---X-ray CBR spectrum by:
\begin{equation}
\Gamma\subHI = 4 \pi ~ \int_{\nu\subHI}^{\infty} d \nu ~ 
\frac{J_{\mathrm{CBR}}(\nu)}{h\nu} ~ a\subHI(\nu) ~ .
\end{equation}
Here, $a\subHI$ is the known cross-section of \HI\ (see Appendix
\ref{ch:rates}).  For a power law spectrum, this integral can be
performed analytically, which makes it possible to fix the
normalisation constant $J_{21}$ for a given spectral index $\eta$.

We consider two fiducial values for $\eta$.  The first, $\eta = 2$,
mimics a typical quasar spectrum \citep{ScottEtAl}; the second, $\eta
= 0.7$ is chosen to match cosmological simulations of the
Lyman-$\alpha$ forest and the observed X-ray CBR up to 40 keV
\citep{HaimanReesLoeb}.  Using the \citet{ScottEtAl} value for
$\Gamma\subHI$, these two choices correspond to values for $J_{21}$ of
0.091 and 0.071, respectively.

\section{Results}
  \label{ch:results}

We determine the amount and distribution of gas (assuming
chemo-dynamical equilibrium at $T = 10^4$ K) that is subject to \htwo\
cooling in the presence of the CBR.  Where the cooling rate due to
\htwo\ processes exceeds the total rate of heating due to the CBR
(\ie\, where the gas is thermally unstable at 10$^4$ K) the gas is
deemed `shielded'.  If this gas is also Toomre unstable at $T \ge 300$
K (\ie\ if it will be gravothermally unstable in the cold phase), then
it is deemed `unstable'; this gas is then prone to star formation.
Additionally, by determining the ISRF intensity required to restore
thermal balance at $T = 10^4$ K at each `unstable' point (so restoring
the situation of gravothermal stability), we estimate the minimum
global SFR for each model.


A few trial models have been recalculated with greater precision to
ensure that numerical effects do not significantly alter our results;
our principal results, shown in Figures \ref{fig:shielding} and
\ref{fig:breakdown}, are convergent to better than 1 percent.

\subsection{Localised Properties of Model Disks
  \label{ch:local}}

\begin{figure}[h]
\centering \includegraphics[height=19.55cm]{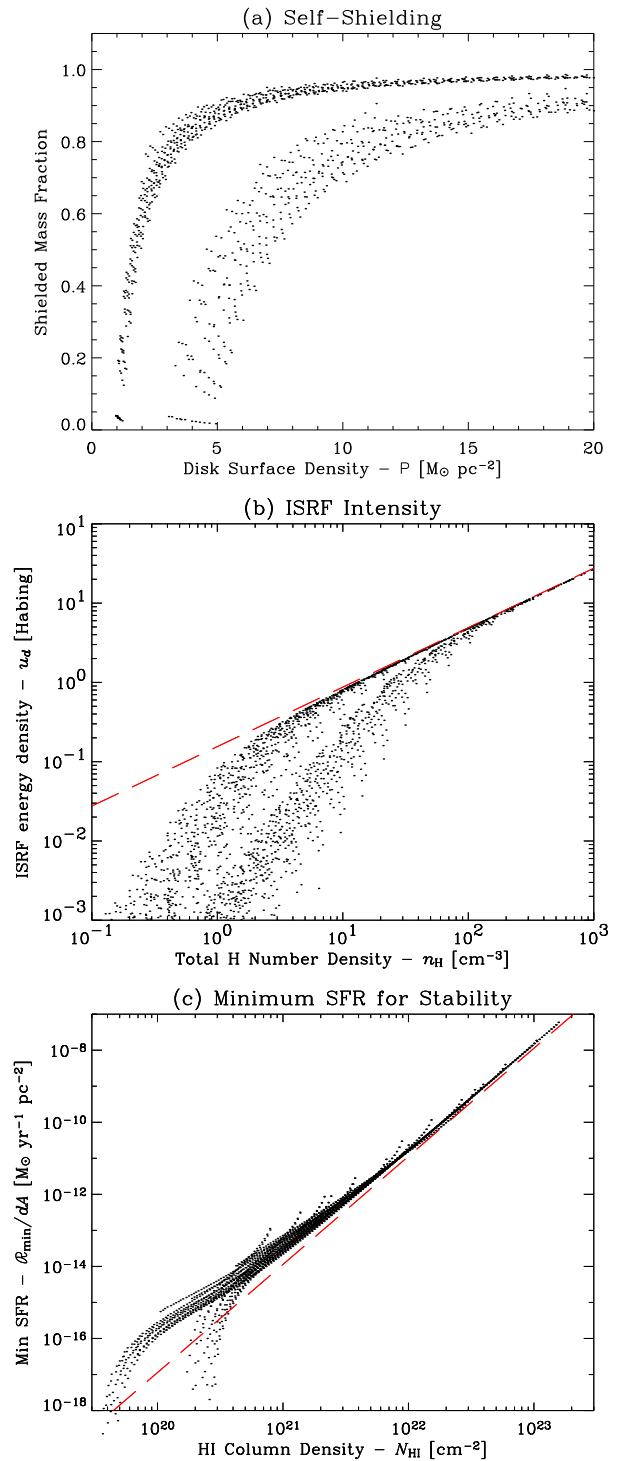}
\caption{\footnotesize{Local scaling relations arising in the model:
    (a) The fraction (by mass) of gas that is shielded from the CBR as
    a function of surface density; (b) the minimum ISRF intensity
    required to maintain thermal balance at $10^4$ K as a function of
    density; (c) the minimum SFR required to maintain thermal balance
    as a function of \HI\ column density.  Representative points are
    drawn from each of the 32 parameter sets we have modelled.  Within
    each panel, the two distinct curves correspond to $\eta = 2.0$
    (upper curves) and $\eta = 0.7$ (lower curves).  {\em Panel} (c)
    {\em should not be interpreted in terms of a Schmidt relation.}
    The SFR plotted in panel (c) is the SFR required elsewhere in the
    cloud to prevent SF in a column of density $N\subHI$; it is not
    the SFR expected in that column.
    \label{fig:local}}}
\end{figure}

In Figure \ref{fig:local}, we present some of the `local' scaling
relations that arise within the models.  Within each panel, the two
distinct curves in regions of marginal shielding correspond to the two
choices of CBR spectral index: the upper curve to $\eta = 2.0$, the
lower one to $\eta = 0.7$.

First, in panel (a), the mass fraction of shielded gas is plotted as a
function of surface density.  These results are in good agreement with
the first estimate made in \textsection\ref{ch:sbcrit}.  A universal
vertical density profile would ensure that all columns with the same
$\Sigma$ would be self-similar; the scatter in this plot ---
particularly in the $\eta = 2.0$ case, which has a higher proportional
X-ray flux --- therefore demonstrates of the importance of detailed
modelling of the vertical gas distribution.  Within the scatter, the
trend with increasing mass is right to left; more massive disks are
better able to self-shield.

\begin{figure*}[t]
\includegraphics[scale = 0.92]{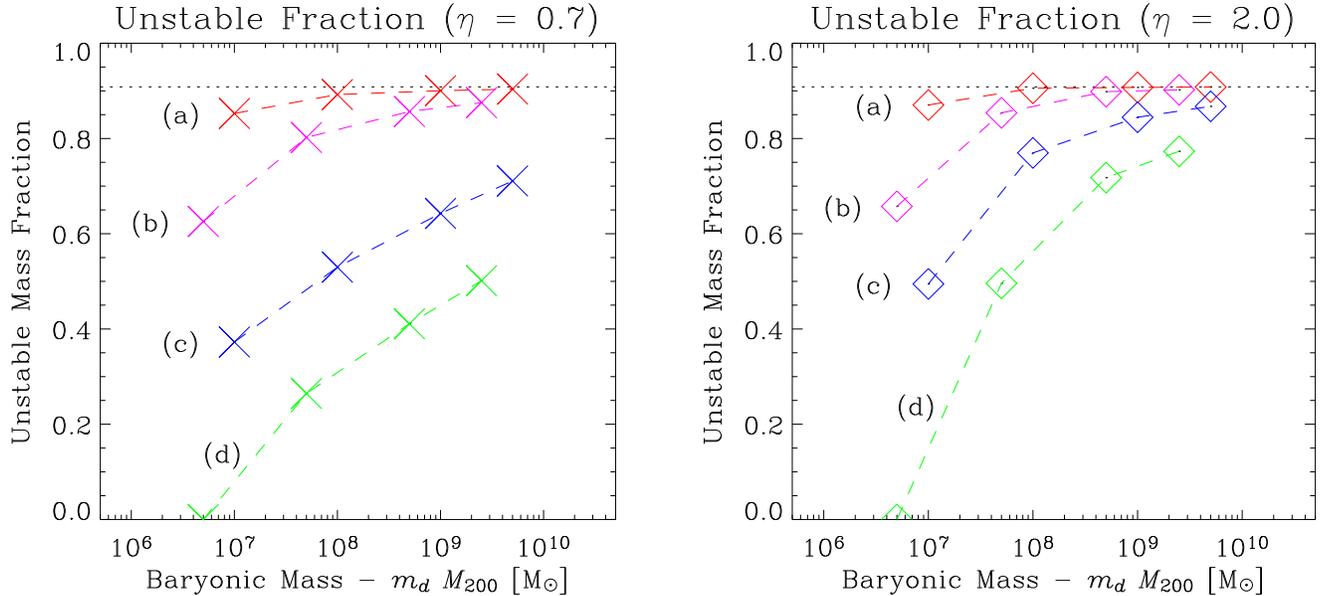}
\caption{\footnotesize{The fractional mass that is unstable to SF in
    the absence of an internal ISRF.  The left panel shows models with
    $\eta = 0.7$; the right shows $\eta = 2.0$.  Within each panel,
    from top to bottom, parameter sets are ($\lambda, ~m_d$) = (0.04,
    0.10), (0.04, 0.05), (0.10, 0.10), \mbox{(0.10, 0.05)}.  In other
    words, lower spin galaxies are more unstable than higher spin;
    then for the same spin, a higher fractional disk mass reduces
    stability.  Note that for an exponential disk, only $\sim 90$
    percent of the mass resides within $4R_d$; this point is shown by
    the dotted line.  Even for the `worst case' CBR, the majority of
    baryonic matter is unstable to SF in at least 50 percent of all
    galaxies with baryonic masses greater than a few times $10^6$
    M\sun ; the same is true for all galaxies with baryonic masses
    greater than $\sim 10^9$ M\sun .
    \label{fig:shielding}}}
\end{figure*}

In panel (b), the minimum ISRF intensity required to maintain thermal
balance at $10^4$ K, $u_d$, is plotted as a function of gas density.
The ISRF is presented in units of the Habing `flux' (the observed
local ISRF energy density at 1000 \AA), $g = 40 \times 10^{-18}$ erg
\AA$^{-1}$ cm$^{-3} = 1.33 \times 10^{-29}$ erg Hz$^{-1}$ cm$^{-3}$
\citep{Habing}.  Within the self-shielded region, $u_d =
0.16~n\subh^{0.75}~g$; this relation has been overplotted.

Finally, in panel (c), the minimum SFR (per unit projected area)
required to preserve thermal balance, $\Rmin/dA$, is plotted as a
function of \HI\ column density.  Again, within the self-shielded
region, the curves converge towards a power law: $\Rmin/dA \approx 1.1 \times
10^{-14} ~ (N\subHI / 10^{21}$ cm$^{-2})^3 = 2.2\times10^{-17} ~
\Sigma\subHI^3$.  The results plotted in this panel agree very well
with those of \citet{CorbelliGalliPalla}.  Note that surrounding the
completely self-shielded region is one in which $\Rmin/dA$ exceeds
this power law: in this region, the CBR actually {\em aids} SF, as the
increased electron abundance due to X-ray ionisation of \HeI\ enhances
\htwo\ production.

Since thermal balance within the self-shielded region is achieved
primarily through photodissociation of \htwo\ (rather than heating by the CBR), the results of the ISRF
calculation are sensitive to the \htwo\ photodissociative flux alone.
The predictions for $u_d$ are therefore robust, in that they should
not be sensitive to the assumed ISRF spectral shape, or to the exact
distribution of the gas.  However, the link between $u_d$ and
$\Rmin/dV$ is more tenuous, since the exact relation depends on the
extent, location, and character of SF within the cloud.

\subsection{Integrated Properties of Model Disks
  \label{ch:global}}

Figure \ref{fig:shielding} shows the fractional mass of gas available
for star formation for each of the 32 parameter sets that we have
modelled.  All else being equal, high spin galaxies are less able to
self-shield than those with lower spin.  In the high spin case, $R_d$
is typically 8---10 times greater than for low spin, which reduces the
surface density at a fixed $R/R_d$ by a factor of order 50---100.

Also, a CBR spectrum with a shallower spectral index has a greater
proportional X-ray flux, and so is able to penetrate deeper into the
cloud.  At a fixed point, this increases the rate of photoheating; at
the same time, the free electrons from helium ionisations lead to an
increased \htwo\ production rate.  Thus, in the $\eta = 0.7$ case,
even though the CBR aids SF near the shielding boundary, this boundary
is pushed deeper within the gas.

Recalling that the adopted values for $\lambda$ represent the 50 and
90 percent points of the expected distribution, the majority of the
disk is predicted to be unstable to SF in at least 50 percent of
putative dark galaxies with baryonic masses greater than $5 \times
10^6$ M\sun .  For baryonic masses greater than $10^8$ M\sun , the
majority of mass is unstable in as much as 90 percent of all
protogalaxies, depending on the distribution of $m_d$ and the true CBR
spectrum.

\begin{figure*}[t] \centering
\includegraphics[scale = 1.]{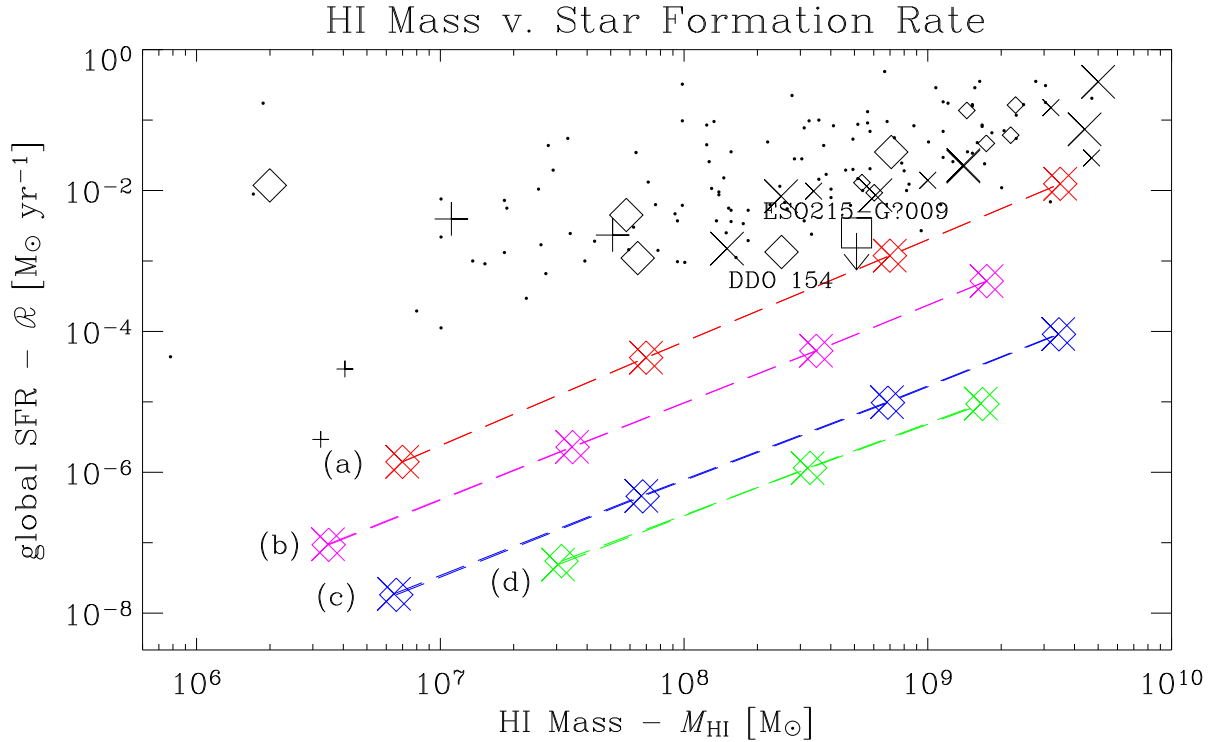}
\caption{\footnotesize{Minimum SFR required for stability, $\Rmin$,
    plotted against \HI\ mass, $M\subHI$.  As in Figure
    \ref{fig:shielding}, like models are connected with a dashed line
    to guide the eye: from top to bottom, models are ($\lambda, ~m_d$)
    = (0.04, 0.10), (0.04, 0.05), (0.10, 0.10), (0.10, 0.05); $\eta =
    0.7$ (2.0) models are plotted with crosses (diamonds).  Note that
    the top and bottom pairs of lines represent the 50 and 90 percent
    points of the expected spin distribution, respectively, and that
    we have deliberately attempted to place a firm lower bound on
    $\Rmin$.  For comparison, observed values of $M\subHI$ and SFR for
    a variety of dwarf galaxies are overplotted, including: `nearby'
    (large pluses) and `distant' (small pluses) dwarfs from
    \citet{HunterElmegreenBaker}; `low' SFR (small crosses) and `high'
    SFR (large crosses) nearby \HI -bright dwarfs from
    \citet{YoungEtAl}; `LSB' (small diamonds) and `normal' (large
    diamonds) dwarfs from \citet{vanZeeEtAl}; 121 members of a sample
    of irregular galaxies (points), spanning a large range in galactic
    parameters, from \citet{HunterElmegreen}; and the extremely \HI
    -rich dwarf ESO215-G?009 (large square)
    \citep{WarrenJerjenKoribalski}. See \textsection\ref{ch:MHI-SFRRel}
    for a discussion of the significance of this relation.
    \label{fig:breakdown}}}
\end{figure*}

This result is directly contrary to that of \citet{VerdeOhJimenez},
who have predicted that all disks in halos with $M_{200} < 10^9$
M\sun\ should remain dark.  However, these authors have assumed a
constant $c_s = \sigma$ = 6 km s$^{-1}$ when calculating the Toomre
parameter, which implies $T \approx 8500$ K; in the context of this
model, this is tantamount to ignoring \htwo\ cooling.

The predicted global minimum SFR within the putative dark galaxies,
$\Rmin$, is plotted in Figure \ref{fig:breakdown}, as a function of
the \HI\ mass, $M\subHI$; in all cases, $M\subHI / m_d \mtwo$ is
between 0.56 and 0.70.  For comparison, some observed values of
$M\subHI$ and SFR for actual dwarf galaxies have been overplotted.  We
have included: a sample of six ``relatively nearby'' ($D \lesssim 15$
Mpc) Im and seven ``more distant'' ($30 \lesssim D \lesssim 100$ Mpc) Im
and Sm galaxies from \citet{HunterElmegreenBaker}; four \HI -bright
nearby galaxies, two each with ``low'' ($< 10^{-4}$ M\sun\ yr$^{-1}$)
and ``high'' ($> 10^{-4}$ M\sun\ yr$^{-1}$) SFRs, from
\citet{YoungEtAl}; seven LSB dwarf galaxies and four ``normal'' dwarfs
from \citet{vanZeeEtAl}; 121 irregular galaxies from the sample of
\citet{HunterElmegreen}, which span more than 8 mag in absolute
magnitude and surface brightness; and ESO215-G?009, an extremely \HI
-rich dwarf galaxy, with the highest accurately measured \HI\
mass:light ratio to date, $M\subHI / L_B \sim 20$
\citep{WarrenJerjenKoribalski}.

For each combination of $\lambda$ and $m_d$, $\Rmin \propto
M\subHI^{1.4}$.  We note that this relation has the same power as the
empirical Schmidt SF law, SFR$/dA \propto \Sigma\subHI^{1.4}$ (Schmidt
1959; Kennicutt 1989, 1998; but see also Boissier et al. 2003).
Moreover, even though these samples were selected to include widely
disparate populations, these populations are not readily distinguished
in Figure \ref{fig:breakdown} for \mbox{$M\subHI \gtrsim 10^8$ M\sun,}
nor is the exceptional ESO215-G?009 clearly differentiated from the
others.

These results are essentially independent of the CBR spectral index.
For the \HI\ fraction, which is determined principally by the
outermost gas in the cloud, this is because the CBR spectrum has been
normalised using $\Gamma\subHI$.  On the other hand, $\Rmin$ is
determined principally by the dense gas closest to the plane of the
disk and to the centre of the galaxy, where the effect of the CBR is
least.  In combination with the results in Figure \ref{fig:shielding},
this gives the surprising result that $\Rmin$ is essentially
independent of the amount of gas that is deemed `unstable'; it depends
on $M\subHI$ alone.

\section{Discussion}
  \label{ch:discussion}

\subsection{The Effects of Substructure and Turbulence
  \label{ch:turbulence}}

It is difficult to be quantitative about the effect of substructure,
but, qualitatively, self-shielding is increased where the surface
density is enhanced.  Since, to first order, the CBR prevents a skin
of fixed column depth from entering the cold phase, removing matter
from a marginally shielded column and adding it to a well shielded
column will increase the overall shielded fraction.  That is, the
omission of substructure such as bars and turbulent cloudlets should
again lead to an underestimate of the amount of gas prone to SF.

\citet{WadaMeurerNorman} have shown that statistically stable
turbulence, leading to a multi-phase interstellar medium, can be
produced and sustained with no stellar energy input.  It is
conceivable that this turbulence could dominate the velocity
dispersion, so providing a third mechanism for support against
gravitational collapse.  The turbulent velocity dispersion required to
support gas at $T \sim 300$ K, however, is typically more than 15 km
s$^{-1}$, whereas in the model these authors present, the turbulent
velocity dispersion is only \mbox{1---5 km s$^{-1}$.}

Moreover, \citet{WadaNorman} have shown that the timescale for cooling
is far shorter than the several dynamical times required to establish
this turbulent motion.  In their model, the central gas cools within
0.1 Myr, whereas turbulent effects first become apparent only after
$\sim$ 2 Myr, and stabilise after $\sim$ 10 Myr.  While turbulent
effects are not unexpected in these galaxies, we therefore expect them
to be insufficient to stabilise the gas, and to emerge only after the
first bout of star formation.

\begin{figure}[t]
\centering
\includegraphics[scale=0.965]{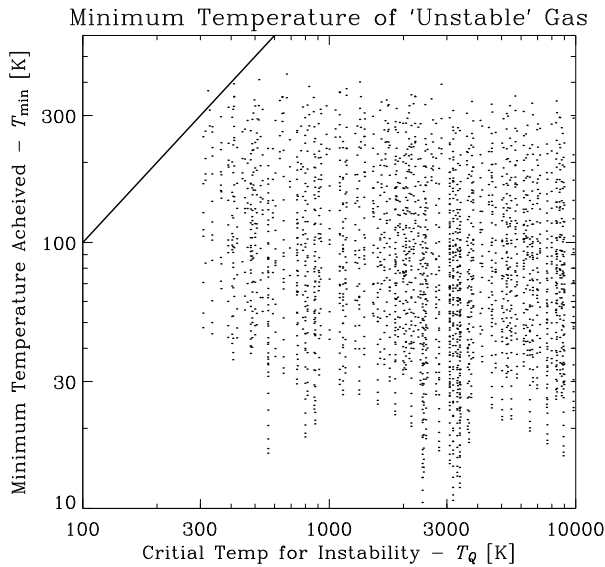}
\caption{\footnotesize{Minimum temperature achieved by cooling gas,
    $T_{\mathrm{min}}$, plotted against the critical temperature below
    which the gas becomes Toomre unstable, $T_Q$, for representative
    points in regions deemed unstable.  Where $T_Q > T_{\mathrm{min}}$,
    gas is able to make an evolutionary transit to a Toomre unstable
    cold phase, in line with the argument presented in
    \textsection\ref{ch:sfcriterion}.  This is essentially always
    true.
    \label{fig:tmin}}}
\end{figure}

\subsection{An Important Sanity Check \label{ch:tmin}}

The crux of our argument is the assumption that, if and when \htwo\
cooling is initiated, the gas will rapidly evolve into a cold phase at
$T \sim 300$ K.  In \textsection\ref{ch:times}, we have argued that
this transition will be fast in comparison to the dynamical timescale;
we are now in a position to verify that this transition will, in fact,
lead to Toomre instability.

Detailed integration of the coupled rate equations encapsulated by
equation (\ref{eq:dxidt}) have found that gas cooling from $T \gtrsim
10^4$ K to $T \lesssim 300$ K develops a non-equilibrium abundance of
\htwo , $x\subhtwo \sim 10^{-3}$ \citep{SnK87,SusaEtAl,OhHaiman},
nearly independently of the initial conditions.  This occurs at $T
\sim 4000$ K, when (two-body) collisional \htwo\ production and
consumption processes become inefficient, leaving an un-replenishing
`freeze-out' abundance of \htwo .  In the presence of a
photodestructive radiation field, these molecules can be quickly
destroyed, preventing further cooling.  \citet{OhHaiman} have shown
that, because of the scalings in the problem, the final temperature
that a parcel of gas (at fixed $n$) is able to cool to depends only on
the ratio $J/n$.

\citet{OhHaiman} describe a method for estimating the minimum
temperature achieved by a cooling parcel of gas, $T_{\mathrm{min}}$, by
finding the temperature for which the timescales for cooling, $t_c$,
becomes longer than that for photodissociation, \mbox{$t_d =
\Gamma\subhtwo^{-1}$}.  With the approximation $\Lambda\subhtwo
\propto T^4$, and assuming the freeze-out abundance, $x\subhtwo =
10^{-3}$, this leads to the following expression for $T_{\mathrm{min}}$:
\begin{equation}
T_{\mathrm{min}} \approx 9000 ~ 
\left(\frac{\beta ~ J_{21}}{n}\right)^{1/3} ~{\mathrm K}~.
%
\end{equation}
(Recall that $\beta$ is the \htwo\ self-shielding factor, discussed in
\textsection\ref{ch:cbr}.)  It should be noted that this analysis does
not include dust catalysed \htwo\ production at $T \lesssim 3000$ K.

In Figure \ref{fig:tmin}, we plot $T_{\mathrm{min}}$ (for gas in
unstable regions) as a function of the critical temperature below
which the gas becomes Toomre unstable, $T_Q$.  Points that lie above
the line $T_{\mathrm{min}} = T_Q$ represent regions where the gas will
transit to a Toomre stable cold phase, in violation of the assumption
on which we base our SF criterion.  There do appear to be very rare
violations of this assumption, however these points are those closest
to the shielding boundary (far from the midplane), in the most Toomre
stable regions (far from the centre) and make an insignificant
contribution to the important integrated quantities.

\subsection{Implications for Star Formation}

Our model argues for a subtle but significant alteration to the (often
unstated) conventional wisdom concerning SF: that \HI\ is the raw
material from which stars form.  Within this paradigm, \HI\ is
commonly referred to as `unprocessed' gas, and it is common and
reasonable to interpret the Schmidt `law' as a {\em causal} relation
between an amount of gas and the level of SF.

The primary role of \htwo\ cooling in this picture lies in the
fragmentation of a collapsing gas cloud into protostellar cloudlets.
We have shown, however, that \htwo\ can also be important in
initiating the first Toomre/Jeans instability.  This is particularly
important in the case of low mass galaxies which would remain Toomre
stable if their thermal evolution were governed by \HI\ alone.

The importance of this slight revision lies in the interpretation of
\HI\ observations.  The picture outlined above interprets \HI\ as, in
some sense, the `original' state of hydrogen.  In the alternate
scenario, however, \htwo\ can be seen as a `default' state for
hydrogen: wherever possible, \HI\ will always develop into \htwo
. Moreover, since \htwo\ cooling virtually guarantees Toomre
instability, some external source of UV radiation is required to
prevent the production of stars from within the gas by dissociating
\htwo\ into \HI .  In this new paradigm, \HI\ is thus a {\em product}
of SF rather than being its primary {\em fuel}, and the Schmidt
relation is {\em symptomatic} of the mode of SF.

In other words, the Schmidt relation should not be interpreted as
`given more \HI , more stars will form', but instead as `more UV
radiation (produced by short-lived O---B stars) is required to support
a bigger \HI\ cloud.'  [For a detailed exposition of this argument,
see \citet{Allen}.]

This revision is not so controversial: the importance of \htwo\
cooling in low mass galaxy formation has been well established by
works such as \citet{PeeblesDicke}, \citet{LeppShull84},
\citet{HaimanThoulLoeb}, and \citet{BabulKepnerSpergel}.  Although
these works are typically (but not exclusively) focused on the effect
that radiative feedback from the first generation of stars has on
galaxy formation, they clearly demonstrate the importance of \htwo\ in
the global thermal and dynamic evolution of galaxies before
protostellar fragmentation.  Closer to home, the established physics
of photodissociation regions [see \citet{HollenbachTielens} for a
comprehensive review] demonstrate the importance of the UV ISRF in
producing \HI\ in our galaxy.

\subsection{The $M\subHI$---SFR Relation \\
and DDO 154--like Galaxies}
\label{ch:MHI-SFRRel}

Our model predicts a hitherto undocumented relation between \HI\ mass
and SFR, which seems to hold over the  range $M\subHI \sim
10^8$---$10^{10}$ M\sun: SFR $\propto M\subHI^{1.4}$.  Within the
model, this relation is a manifestation of self-regulating SF
\citep{LinMurray, OmukaiNishi}, in which the ISM is kept warm and
stable by a UV ISRF that is maintained by constant regeneration of
O---B stars.

The slope of the relation is driven by the distribution of the densest
gas in galaxy centres, since it is this gas that requires the most
radiation to maintain thermal stability.  The scatter is set primarily
by the distribution of galactic spins, as well as variations in the
relative mass of the disk in comparison to the halo, since, for a
fixed mass, these determine the density near the centre.

Because the model was designed to place a lower limit on the SFR, it
cannot predict the zero-point of such a relation.  Within the
self-shielded region, the \htwo\ cooling rate is typically less than
1---10 percent of the net cooling rate; a significantly higher ISRF
might be required to balance these other cooling mechanisms.
Moreover, if there were, for example, a systematic run with mass in
the relative contribution of metal-line cooling in comparison to
\htwo\ cooling, the inclusion of metals might alter the predicted
slope of the relation.  We stress that this would only be true,
however, if metal-line cooling were important in {\em initiating} the
transition from warm to cold.

The general agreement between the character of the predicted and
observed $M\subHI$---SFR relations lends weight to the physical
picture on which this simple model is based: the results in Figure
\ref{fig:breakdown} suggest that the well observed population of dwarf
galaxies represent the minimum rate of ISRF-regulated SF in galaxies.

Moreover, in the range $M\subHI \sim 10^8$---$10^{10}$ M\sun, it seems
that the majority of dwarfs are forming stars in the same way:
`extreme' objects like DDO 154 and ESO215-G?009 are not distinguished
from `normal' dwarfs in the $M\subHI$---SFR plane.  In response to the
question ``Why was so little gas processed into stars?", therefore,
these galaxies' high \HI\ mass:light ratios are not due to
inefficiencies in the present mode of SF.  Instead, they may be, for
example, a reflection of late formation times or differing
environmental effects.

\subsection{The Complete Prevention of SF}

Figures \ref{fig:shielding} and \ref{fig:breakdown} show that SF is
completely prevented in only one of the models:
($\mtwo,~\lambda,~m_d$) = ($10^8$ M\sun , 0.10, 0.05).  As is apparent
in Figure \ref{fig:toomre}, this is due primarily to rotational
support rather than an inability to self-shield: the gas is
self-shielding at $R \approx$ 1 (2.5) $R_d$ for $\eta = 0.7$ (2.0),
but is never Toomre unstable at \mbox{$T = 300$ K}.  However, if the
gas were able to cool to \mbox{$\sim 30$ K} (\ie\ if it could initiate
metal line cooling), then up to 70 percent of the gas would become
available for SF, and we would predict $\Rmin \sim 10^{-9}$ M\sun\
yr$^{-1}$.

\subsection{The Detectability of Dark Galaxies
  \label{ch:detection}}

Throughout this work, since the focus has been on gravothermal
stability in the absence of internal SF, it has been implicitly
assumed that the gas is predominantly \HI\ (and not H{\footnotesize
II} or \htwo).  Clearly, a very large stellar component would be
necessary to completely ionise the gas, and it has been shown that the
gas becomes gravothermally unstable in the presence of small amounts
of \htwo .  It has also been argued that, once SF begins, the ISRF
will regulate the SFR by dissociating \htwo\ into \HI .

\HI\ is detectable by the hyperfine transition at $\sim 21$ cm,
provided that the excitation temperature, $T_{\mathrm{spin}}$, exceeds
the mean background radiation temperature, $T_{\mathrm{bg}} \approx 3$K.
$T_{\mathrm{spin}}$ can be written \citep{HIarticle}:
\begin{equation}
T_{\mathrm{spin}} = \frac{T + y~T_{\mathrm{bg}}}{1+y}~,
\end{equation}
where, neglecting \HI\ excitation by electron collisions, $y \sim (T /
1000$ K) / ($n\subHI / 0.2$ cm$^{-3}$).  From Figure \ref{fig:local},
it is apparent that $n\subHI$ is essentially always in the range
10$^{-1}$---10$^2$ cm$^{-3}$ within the self-shielded region, in which
case $T_{\mathrm{spin}}$ = 800---9800 K.  Moreover, since \HI\ is so
optically thick to Lyman-$\alpha$, even the low predicted SFRs act to
thermalise $T_{\mathrm{spin}}$ via scattered Lyman-$\alpha$ photons.

Without exception, all models are thus predicted to have detectable
\HI\ emission, independently of their optical properties.  Joint
optical--plus--21cm surveys are therefore predicted to yield complete
samples of galaxies.

\section{The Existence of Dark Galaxies --- Summary}
\label{ch:summary}

We have developed a model for the long-time configuration of a
baryonic dark galaxy, in order to determine whether such a galaxy,
having formed, could conceivably remain dark.  The model has the
following features:

\begin{itemize}

\item The gas is in a dynamically equilibrated disk (rotationally
  supported in the radial direction and in hydrostatic equilibrium
  vertically) at the end of the galaxy formation process.

\item The gas is isothermal at $T = 10^4$ K, which represents the
  endpoint of its thermal evolution in the absence of \htwo\ cooling.
  Together, the assumptions listed above limit the model to
  considering galaxies with $\mtwo \gtrsim 10^8$ M\sun .

\item The gas is in chemical equilibrium, in the presence of a
  UV---X-ray CBR.  The CBR can prevent SF by preventing the gas from
  undergoing a transition (driven by \htwo\ cooling) from a warm,
  largely stable phase to a cold, mostly unstable phase.

\end{itemize}

We evaluate the stability of this situation of thermal, chemical and
  dynamical equilibrium by determining whether or not \htwo\ cooling
  can induce gravothermal instability, and hence SF, in the face of
  the CBR.  Additionally, wherever an ISRF is required (in addition to
  the CBR) to counterbalance \htwo\ cooling, we place an approximate
  lower bound on the expected SFR.

We find that, in the absence of an internal UV radiation field, gas
that is sufficiently self-shielded to support an \htwo\ abundance as
low as 10$^{-4}$ is unable to remain gravothermally stable.  This
statement remains true independently of any other cooling processes,
including metal line cooling: so as to place a firm lower bound on the
amount of gas that will become gravothermally unstable, we have
ignored the significant shielding and cooling effects of dust and
metals when calculating the \htwo\ cooling rate at 10$^4$ K.
Moreover, if the gas is even mildly enriched, then its final
temperature drops by an order of magnitude from $\sim$ 300 K to $\sim$
30 K \citep{NormanSpaans}; particularly for the models most affected
by the CBR, this could significantly reduce stability against SF
(Figure \ref{fig:toomre}).

Our conclusions can be summarised as follows:

\begin{enumerate}

\item The model predicts a correlation between \HI\ mass and SFR,
  which is observed in a wide variety of dwarf galaxies (see Figure
  \ref{fig:breakdown}).  This suggests that dwarf galaxies represent 
  the minimum levels of ISRF-regulated SF in the universe.  Further 
  theoretical and observational studies of this relation should prove 
  fruitful.

\item With only one exception, the modelled galaxies cannot avoid SF
  indefinitely, and so will develop a stellar component.  Even for
  baryonic masses as low $5 \times 10^6$ M\sun , the majority of gas
  in greater than 50 percent of all halos is unstable to SF in the
  absence of an internal ISRF (Figure \ref{fig:shielding}).

\item Without exception, all models would have detectable \HI\
  emission.  That is, whatever dark or dim galaxies exist should
  be detectable in \HI PASS--class surveys.

\end{enumerate}

Above the detection limit of $M\subHI \sim 7 \times 10^6$ M\sun\
\citep{MeyerEtAl}, \HI PASS did not detect any extragalactic \HI\
clouds in the Local Group that did not also have a stellar component.
Similar results are available for other groups and clusters
\citep{Zwaan, Waugh}.  Within these limits, there is no reason to
suspect a population of undetected baryonic dark galaxies.

\vspace{0.6 cm}

We wish to thank Stuart Wyithe for many fruitful discussions in the
course of this work, and Joop Schaye for his helpful and insightful
responses to an early draft.  We also thank the anonymous referee,
whose comments led to the calculation presented in \textsection\ref{ch:tmin}.
ENT was supported by a University of Melbourne--CSIRO collaborative
grant while this work was completed.

\appendix

\section{Chemical Processes in the Disk} \label{ch:rates}

The full list of reactions considered in this study is given in Table
A1, along with their rate coefficients.  All quantities in this
section are listed in cgs units.  The symbol $T_n$ is used to denote
$T^{-n}$.  Note that the rate coefficient for reaction (21) has two
terms; the second term arising from dielectric recombination
\citep{Cen92}.  The rates given for reactions (16) and (17) both are
valid in the low density ($n_H < 10^{4}~ \mathrm{cm^{-3}}$) limit.

Also note that three of the expressions given in Table 3A of
\citet{HaimanThoulLoeb} differ from those given in the works that
those authors cite.  They give the exponent of $T_3$ in the expression
for the rate coefficient of reaction (2) as 0.2; it should read -0.2
[This typographical error is not present in \citet{HaimanReesLoeb}].
More seriously, the multiplying factor in the expression for reaction
rate (3), given as 5.57, should read 3.23.  Also, the exponent of $T$
in reaction rate (6), which should be 2.17, is omitted in
\citet{HaimanThoulLoeb}.  These alterations reduce the equilibrium
abundances of \htwo\ by several orders.

\begin{table}[h] \begin{center}
\begin{tabular}{ r r l l r }
\hline
\hline
\multicolumn{5}{c}{ } \\
\#&   
\multicolumn{2}{c}{Reaction}   &
Rate Coefficient, $\alpha$ (cm$^3$ s $^{-1}$)&  
Ref.
\\
\hline
(1) &
\hz + \eminus \goesto & \hp + 2\eminus
&$
5.85\times10^{-11} ~ T^{1/2} ~ e^{-157809.1/T}~(1 + T_5^{1/2})^{-1}
$&
Cen92
\\
(2) &
\hp + \eminus \goesto & \hz + $h\nu$
&$
8.40\times10^{-11} ~ T^{-1/2} ~ T_3^{-0.2} ~ (1 + T_6^{0.7})^{-1}
$&
Cen92
\\
(3) &
\hz + \eminus \goesto & \hm + $h\nu$
&$
3.23\times10^{-17}~T^{1/2}
$&
HTL96
\\
(4)&
\htz + \eminus \goesto & \hm + \hz
&$
2.7 \times10^{-8} ~ T^{-3/2} ~ e^{-43000/T}
$&
HTL96
\\
(5) &
\hz + \hm \goesto & \htz + \eminus
&$
1.3\times10^{-9}
$&
Raw88
\\
(6) &
\hz + \hm \goesto & 2\hz + \eminus
&$
5.3\times10^{-20} ~ T^{2.17} ~ e^{-8750/T}
$&
S\&K87
\\
(7) &
\eminus + \hm \goesto & \hz + 2\eminus
&$
4 \times 10^{-12} ~ T ~ e^{-8750/T}
$&
S\&K87
\\
(8) &
\hp + \hm \goesto & 2\hz
&$
7 \times 10^{-7} ~ T^{-1/2}
$&
D\&L87
\\
(9) &
\hp + \hm \goesto & \htp + \eminus &
$ 4\times10^{4} ~  T^{-1.4} ~ e^{-15100/T} ~ : ~ T > 10^4$ K & \\ &&&
$ 10^{-8} ~ T^{-0.4} ~ : ~T \le 10^4$ K &
S\&K87 
\\
(10) &
\hp + \hz \goesto & \htp + $h\nu$
&
$3.255\times10^{-20}~T -4.152\times10^{-25} ~ T^2 - 6.157e^{-17}
~ : ~ T > 4000$ K & \\ &&&
$1.38\times10^{-23}~T^{1.845} ~ : ~ 200 \le T \le 4000$ K 
&
RDB93
\\
(11) &
\hp + \htz \goesto & \htp + \hz
&$
2.4\times10^{-9}~e^{-21200/T}
$&
S\&K87
\\
(12) &
\hz + \htp \goesto & \htz + \hp
&$
6.4 \times 10^{-10}
$&
HTL96
\\
(13) &
\hm + \htp \goesto & \htz + \hz
&$
5 \times 10^{-6} ~ T^{-1/2}
$&
D\&L87
\\
(14) &
\eminus + \htp \goesto & 2\hz
&$
1.68 \times 10^{-8} ~ (T / 300)^{-0.29}
$&
HTL96
\\
(15) &
\htz + \hz \goesto & 3\hz
&
$6.11\times10^{-14} ~ e^{-29300/T} ~ : ~ T > 7390$ K & \\ &&
&
$2.67\times10^{-15} ~ e^{-(6750/T)^2} ~ : ~ T \le 7390$ K 
&
L\&S83
\\
(16) &
\htz + \htz \goesto & \htz + 2\hz
&$
5.22\times10^{-14} ~ e^{32200/T} ~ : ~ T > 7291$ K & \\ &&
&$ 3.17\times10^{-15} ~ e^{-(4060/T)-(7500/T)^2} : T \le 7291$ K 
&
S\&K87
\\
(17) &
\htz + \eminus \goesto & 2\hz + \eminus
&$
4.38\times10^{-10} ~ T^{0.35} ~ e^{-102000/T}
$&
S\&K87
\\
(18) &
\hez + \eminus \goesto & \hep + 2\eminus
&$
2.38\times10^{-11} ~ T^{1/2} ~ e^{-285335.4/T}~(1 + T_5^{1/2})^{-1}
$&
HTL96
\\
(19) &
\hep + \eminus \goesto & \hepp + 2\eminus
&$
5.68\times10^{-12} ~ T^{1/2} ~ e^{-631515.0/T}~(1 + T_5^{1/2})^{-1}
$&
Cen92
\\
(20) &
\hep + \eminus \goesto & \hez + $h\nu$
&$
1.50\times10^{-10} ~ T^{-0.6353}$ & \\&&&
$~+~1.9 \times10^{-3} ~ T^{-3/2} ~ e^{-470000/T} (1 + 0.3e^{-94000/T})
$&
Cen92
\\
(21) &
\hepp + \eminus \goesto & \hep + $h\nu$
&$
3.36\times10^{-10} ~ T^{-1/2} ~ T_3^{-0.2} (1 + T_6^{0.7})^{-1}
$&
Cen92
\\
\multicolumn{5}{c}{ } \\
\hline
\hline
\end{tabular} 
\end{center}
\textbf{Table A1} --- Reaction rate coefficients.  The expressions
  given for reactions (16) and (17) are valid for $n_H < 10^4~
  \mathrm{cm^{-3}}$.  References: (Cen92) \citet{Cen92}; (D\&L87)
  \citet{DnL87}; (HTL96) \citet{HaimanThoulLoeb}; (L\&S83)
  \citet{LnS83}; (Raw88) \citet{Raw88}; (RDB93) \citet{RDB93};
  (S\&K87) \citet{SnK87}.  All coefficients are given in cgs units.
  Note that the rate coefficients for reactions (2), (3) and (6) given
  in \citet{HaimanThoulLoeb} disagree with those given in the works
  those authors cite (see text).
\end{table}

For photoionisation and photodissociation reactions, we employ the
reaction cross-sections given by \citet{AbelEtAl}, except for
ionisation of \hez\ and \htz, for which we use the more recent
expressions given by \citet{YanSadeghpourDalgarno}.  The different
asymptotic behaviour of the expression for \hez\ ionisation has a
significant effect on the chemical structure of our models, but does
not drastically affect our conclusions.  When calculating the
cross-section for \htwo\ ionisation, we assume all \htwo\ to be in the
para- configuration.  This is appropriate only when reactions (5) and
(12) are the dominant \htwo\ production mechanisms; within the model
these reaction rates are typically seven orders of magnitude higher
than for reaction (13).

To account for self-shielding of \htwo\ against LW photons, including
the effects of Doppler broadening and line overlap, we use the
shielding factor of \citet{deJongDalgarnoBoland}:
\begin{displaymath}
\beta(\tau)= \left(
\left(\frac{\tau^2}{2}~\ln\left[\frac{\tau^2}{\pi}\right]\right)^{-1/2}
+ \left(\frac{\tilde{a}}{\tau}\right)^{1/2}\right) ~ \times
\mathrm{erfc} \left[ \left(\frac{\tau \tilde{a}}{\pi a_{\mathrm H_2^0}}\right)^{1/2}\right] ~ ,
\end{displaymath}
where erfc denotes the complimentary error function, and $\tilde{a} =
9.2 \times 10^{-3} ~ \sigma^{-1}$ cm km s$^{-1}$ is called the Voigt
parameter.  Following \citet{deJong}, we assume that each Solomon
dissociation heats the gas by 2.5eV.

Finally, we use the lengthy expression given by \citet{LnS83} for the
rate of \htwo\ cooling, $\Lambda_{\txth_2}$:

\begin{eqnarray}
\Lambda_{\mathrm{H}_2} = \left(\frac{L_{vH}}{1 + (L_{vH}/L_{vL})} +
  \frac{L_{rH}}{1 + (L_{rH}/L_{rL})}\right) ~n_{\mathrm{H}_2}~~ : ~ 
&&  L_{vH} = 1.10\times10^{-18} ~ e^{-6744/T} \nonumber \\
&& L_{vL} = 8.18\times10^{-13} ~ \left(\alpha_{16} ~ n{_{\mathrm{H}^0}} 
+ \alpha_{17} ~ n_{\txth_2^0} \right) \nonumber \\
&& L_{rH} = 3.90\times10^{-19} ~ e^{-6118/T} \nonumber \\
&& L_{rL} = \left(1.38\times10^{-22} ~ e^{-9243/T}\right) ~ Q_n \nonumber \\
&& Q_n = 1.2 ~ (n{_{\mathrm{H}^0}})^{0.77} + (n_{\mathrm{H}_2^0})^{0.77} ~ , 
\nonumber \end{eqnarray}

\noindent where each of the $L$s has units of ergs s$^{-1}$.  As given here,
this expression is valid only for temperatures above 4031 K.  For the
calculation presented in \textsection\ref{ch:tmin}, we use:
$\Lambda\subhtwo \approx 4 \times 10^{-39} ~ T^4$,
which is a fit given by \citet{OmukaiNishi}, and is a good
approximation for $600 \lesssim T \lesssim 3000$ K and $n \lesssim
10^4$ cm$^{-3}$.  For all other cooling processes, we use the
expressions given in \citet{HaimanThoulLoeb}, and references therein.

\section{Calculating the Quantity $\xi$}
  \label{ch:xicalc}

The quantity $\xi(R,~Z)$ in equation (\ref{eqn:ode}) is related to
what \citet{Bahcall} calls the effective halo potential, except that
any vertical variation in $\xi$ is neglected in that work.  The effect
of explicitly including $\xi$ in determining the volume density of the
disk gas is illustrated in Figure \ref{fig:xi}.  In this section, we
motivate and describe the method by which we have calculated $\xi$ in
our integration of equation (\ref{eqn:ode}).

Throughout this section, we use the symbols $R$ and $Z$ to represent
the (cylindrical) galactocentric radius and the vertical distance from
the midplane, respectively; we reserve the lowercase $r$ for spherical
coordinates.  We will also use the symbols $M,~\rho$, and $V$ here to
refer to quantities which can be expressed analytically, as distinct
from $\emm$, $\varrho$, and $\vee$, which must be computed
analytically.

As described in \textsection\ref{ch:scaleheight}, $\xi$ is defined as:
\begin{equation}
\xi(R,Z) \equiv \frac{1}{4\pi G \rho_0}
\left(4\pi G \varrho_h(R,Z) - 
\frac{1}{R} \frac{\partial \vee ^2(R,~Z)}{\partial R} \right)~.
\label{xidefined} \end{equation}
Here, $\vee ^2(R,~Z)$ is a sum in quadrature of contributions due to
the (spherical) dark matter halo and the (cylindrical) baryonic disk,
\viz\ \citep{Freeman, BinneyTremaine}:
\begin{eqnarray}
\vee^2(R,~Z) && = ~~ \vee_h^2(r) ~~ + ~~ V_d^2(R) \nonumber \\
&&= \frac{G \emm_h(r)}{r} + \frac{G m_d M_{200}}{R_d} ~ 2 y^2
\bigg(I_0(y) K_0(y) - I_1(y) K_1(y) \bigg) ~~ 
: ~ y \equiv \frac{R}{2R_d} ~ ,
\label{rotation}\end{eqnarray}
where $I_n$ and $K_n$ are $n$th order modified Bessel functions of the
first and second kinds, respectively.  Note that there is an implicit
thin disk assumption embedded in this expression, since it neglects
any variation of $V_d$ with $Z$.

The final halo mass distribution, $\emm_h(r)$, 
is determined using the method described by MMW, which assumes that
the `adiabatic invariant', $r M(r)$ \citep{BarnesWhite}, is conserved
throughout the galaxy formation process.  That is, a test particle
initially located at a mean radius $r_i$ moves through successive
radii $r$ in such a way that, at any given time:
\begin{equation}
r_i ~ M_i(r_i) = r ~ \emm (r) = r ~ [ \emm_h(r) + M_d(R) ] ~ .
\label{adiabaticscheme}\end{equation}
Here $M_i(r_i)$ is the (assumed) initial dark-plus-baryonic mass
profile, $M_d(R)$ is the (trial) final disk mass profile, and
$\emm_h(r)$ is the (unknown) adiabatically contracted halo profile,
\viz :
\begin{equation}
{\cal M}_h(r) = (1 - m_d) ~ M_i(r_i) ~ .
\label{finalmass} \end{equation}
Combining equations (\ref{adiabaticscheme}) and (\ref{finalmass}),
we obtain an expression which we solve numerically for $r_i$ as a
function of $r$:
\begin{equation}
M_d(r) + \left(1 - m_d - \frac{r_i}{r}\right) ~ M_i(r_i) = 0 ~ .
\label{rinit}\end{equation}
The resultant $r_i$ can then substituted back into equation
(\ref{finalmass}) and thence (\ref{rotation}) to determine $\vee$.

Of course, determining $\partial \vee^2 / \partial r$ is doubly
expensive, since this numerical procedure must be repeated twice in
order to estimate the derivative; the same is true of $\varrho_h$:
\begin{equation}
\varrho_h(r) = \frac{1}{4 \pi r^2} ~ \frac{d \emm_h(r)}{dr} ~ .
\label{varrho} \end{equation}
However, we can rewrite the halo contribution to the $\partial \vee^2
/ \partial r$ term in equation (\ref{xidefined}) as
\citep{KuijkenGilmore}:
\begin{eqnarray}
\frac{1}{R}~\frac{\partial \vee_h^2(r)}{\partial R} &&=
\frac{1}{R}~\frac{\partial r}{\partial R}~\frac{\partial}{\partial r}
\left[\frac{G {\cal M}_h(r)}{r}\right] \nonumber \\
&&= \frac{1}{R}~\frac{R}{r} ~ \left(\frac{G}{r} ~ 
\frac{\partial {\cal M}_h(r)}{\partial r} - 
\frac{G \emm_h(r)}{r^2}\right)
\nonumber \\  
&&= \frac{1}{r} ~ \left(\frac{G}{r}~4 \pi ~ r^2 ~ \varrho_h(r)
- (1 - m_d)~\frac{G M_i(r_i)}{r^2}\right) \nonumber \\
&&= 4 \pi G ~ \varrho_h(r) - (1-m_d)\frac{G M_i(r_i)}{r^3} ~ ,
\end{eqnarray}
where we have used equations (\ref{varrho}) and (\ref{finalmass}) to
get from the second to the third line.

When this expression is substituted back into
equation (\ref{xidefined}), the $\varrho_h$ terms cancel, and we are
left with:
\begin{equation}
\xi(R, Z) 
= \frac{1}{4 \pi G \rho_0} ~ 
\left((1-m_d)\frac{G M_h(r_i)}{r^3} - \frac{1}{R} ~ 
\frac{\partial V_d^2(R)}{\partial R} \right) ~ .
\end{equation}
In this new expression, both $M_h$ and ${\partial V_d^2}/{\partial R}$
\citep{BinneyTremaine} are known analytically.

Thus, by incorporating a lookup table for $r_i(r)$, we have developed
a means of retaining all contributing terms in equation
(\ref{eqn:ode}), without significantly increasing computation time.
Moreover, with our rewritten definition of $\xi$, we have eliminated
the need to compute $d \emm_h/dr$ and $\partial \vee^2 / \partial r$
numerically.  We found these modifications to decrease the computation
time for the vertical distribution of the gas by a factor of roughly
one hundred.

\end{document}